\documentclass{aa}
\usepackage[T1]{fontenc}
\makeatletter

\providecommand{\LyX}{L\kern-.1667em\lower.25em\hbox{Y}\kern-.125emX\@}

\usepackage[T1]{fontenc}
\makeatletter

\usepackage{epsfig}

\makeatother

\makeatother

\begin{document}

\title{The environments and ages of extragalactic radio sources
inferred from multi-frequency radio maps}

\titlerunning{Environments and ages of radio sources}

\author{C. R. Kaiser}
\institute{Max-Planck-Institut f{\"u}r Astrophysik, Postfach 1317,
Karl-Schwarzschild-Str.1, D-85741 Garching, GERMANY}
\offprints{ckaiser@mpa-garching.mpg.de}

\date{Received 1.5.2000 / Accepted <date>}
\thesaurus{3(11.01.2; 11.09.1 Cygnus A; 11.09.1 3C 219; 11.10.1;
11.17.4 3C 215; 13.18.1)}

\maketitle

\begin{abstract}
A 3-dimensional model of the synchrotron emissivity of the cocoons of
powerful extragalactic radio sources of type FRII is constructed. Bulk
backflow and energy losses of the relativistic electrons, radiative
and adiabatic, are self-consistently taken into account. Thus the
model is an extension of spectral aging methods including the
underlying source dynamics into the age estimates. The discrepancies
between spectral ages and dynamical ages arising from earlier methods
are resolved. It is also shown that diffusion of relativistic
particles within the cocoon is unlikely to significantly change the
particle spectrum and thus the emitted radio spectrum. Projection
along the line of sight yields 1 or 2-dimensional surface brightness
distributions which can be compared with observations. From the model
parameters constraints on the source age, the density of the source
environment and the angle to the line of sight can be
derived. Application of the method to Cygnus A, 3C 219 and 3C 215 show
that the method provides robust estimates for the model parameters for
sources with comparatively regularly shaped radio lobes. The
resolution of the radio maps required is only moderate. Within the
large uncertainties for the orientation angle, the three example
sources are found to be consistent with orientation-based unification
schemes of radio-loud AGN. In the case of Cygnus A the gas density of
the environment is found to agree with independent X-ray
measurements. For 3C 219 and 3C 215 the densities derived from the
model are apparently too low. It is suggested that these discrepancies
are caused by overestimates of slope and core radius of $\beta$-models
for the gas density distribution from X-ray observations for clusters
hosting powerful radio sources.  
\end{abstract}

\section{Introduction}

The `standard model' of extragalactic classical double radio sources
by Scheuer (1974)\nocite{ps74} explains these sources as twin jets
emerging from the Active Galactic Nucleus (AGN) and impinging on the
surrounding gas. The compact, high surface brightness regions or hot
spots at the end of the jets are interpreted as the sites of the
interaction between the jets and the environment. After passing
through the hot spot region, the jet material inflates the radio lobes
or cocoon which is observed as diffuse emission in between the hot
spots and the source core (e.g. Muxlow \& Garrington 1991\nocite{mg91}
and references therein). This picture forms the basis of virtually all
more recent attempts at modeling the dynamics and radio emission
properties of powerful radio galaxies and radio-loud quasars (Begelman
\& Cioffi 1989\nocite{bc89}, Falle 1991\nocite{sf91}, Daly
1994\nocite{rd94}, Nath 1995\nocite{bn95}, Kaiser \& Alexander
1997\nocite{ka96b}, Chy{\.z}y 1997\nocite{kc97}, Kaiser et
al. 1997\nocite{kda97a}, Blundell et al. 1999\nocite{brw99}).

These models for the evolution of individual classical doubles or
FRII-type objects (Fanaroff \& Riley 1974\nocite{fr74}) can be used to
study the cosmological evolution of the population as a whole. One of
the more important trends is the apparent decrease of the mean linear
size of the radio structures with increasing redshift (see Neeser et
al. 1995\nocite{ner95} and references therein).

Two recent attempts in fully explaining the linear size -- redshift
anti-correlation were presented by Blundell et
al. (1999)\nocite{brw99} and Kaiser \& Alexander
(1999a\nocite{ka98a}). Blundell et al. argue that a specific form of
pre-aging of the relativistic particle population in the hot spots in
connection with the lower flux limit of complete observed samples
causes this anti-correlation (the so-called youth-redshift degeneracy,
Blundell et al. 1999\nocite{brw99}, Blundell \& Rawlings
1999\nocite{br99}). Alternatively, Kaiser \& Alexander propose that
the apparent smaller sizes of radio sources at high redshift could be
caused by a denser environment of these objects at high
redshift. Clearly, the evolution or non-evolution of the radio source
environments is of great interest to decide this and other important
cosmological questions.

Various methods have been employed in determining the properties of
the radio source environments. X-ray observations of the hot gas
around radio sources in clusters yield direct estimates of the gas
density (Crawford et al. 1999\nocite{clfbh99}, Hardcastle \& Worrall
1999\nocite{hw99}). However, such studies are at present confined to
objects at low redshifts. Furthermore, the AGN (e.g. Sambruna et
al. 1999\nocite{sem99}) and the large-scale radio structure contribute
to the X-ray emission (Brunetti et al. 1999\nocite{bcsf99}, Kaiser \&
Alexander 1999b\nocite{ka98b}, Sect. \ref{sec:disc}). The properties
of the density distribution inferred from X-ray observations may
therefore be considerably influenced by the presence of the radio
source itself.

Faraday rotation of the polarisation angle of the synchrotron emission
and the related depolarisation can be used to determine the gas
density of the material surrounding the radio lobes (e.g. Garrington
et al. 1988\nocite{glcl88}, Laing 1988\nocite{rl88}). Unfortunately,
this method does not provide a direct measure of the gas density as
the rotation measure also depends on the strength of the magnetic
field in the source environment. Usually it is not possible to break
this degeneracy because the strength of the magnetic field is only
poorly known.

Constraints on the density of the radio source environments also come
from optical or infrared galaxy counts around the host galaxies
(e.g. Hill \& Lilly 1991\nocite{hl91}). It is not straightforward to
decide whether galaxies in the field of the radio source host are
associated or chance background objects. Resolving the ambiguity would
ideally require the spectroscopic measurement of the redshifts of all
objects in question. This is very time consuming. In any case, the
method provides only indirect constraints on the gas density around
the radio source as this has to be inferred from comparison with low
redshift clusters or groups of similar richness.

The ages of powerful radio sources can in principle be determined from
their radio spectrum (e.g. Alexander \& Leahy 1987\nocite{al87}). The
various energy losses of the relativistic particles depend on time and
so the shape of the radio spectrum contains an encoded history of the
source. In practice the time-dependence of the energy losses
complicates the estimation of the spectral age because different parts
of the source have different ages. Even when radio maps at various
frequencies which fully resolve the radio lobes are used, it is
difficult to disentangle the effects of the various loss mechanisms
and possible bulk backflow of the cocoon material along the jet
(e.g. Rudnick et al. 1994\nocite{rka94}).

The model developed in this paper aims at tracing the individual
evolution of parts of the cocoon and thereby providing more accurate
estimates for the source age. At the same time the model also
constrains the density in the source environment and other parameters
like the energy transport rate of the jets. It is solely based on
radio observations which are available for a large number of objects,
even at high redshift. This model may therefore provide an important
step in determining the cosmological evolution of the FRII radio
source population.

In Sect. \ref{sec:spec} I show that diffusion of relativistic
particles in the cocoons of FRII-type objects should not significantly
change the energy distribution of particles.  In Sect. \ref{sec:mod}
the dynamical model of Kaiser \& Alexander (1997, hereafter
KA)\nocite{ka96b} and its extension by Kaiser et
al. (1997\nocite{kda97a}, hereafter KDA) to include synchrotron
emission are briefly summarised. A 3-dimensional model of the
synchrotron emissivity of the cocoon based on this analysis is
constructed in Sects. \ref{sec:spat} and \ref{sec:back}. Methods for
comparing the model predictions with observations are developed in
Sect. \ref{sec:compa}. The degeneracy of model parameters resulting
from the comparison method is also discussed here. The model is then
applied to three FRII-type radio sources, Cygnus A, 3C 219 and 3C 215,
in Sect. \ref{sec:appl}. The results are discussed in Sect.
\ref{sec:disc}. The main conclusions are summarised in Sect.
\ref{sec:conc}.

Throughout this paper I use $H_{\rm o}=50$ km s$^{-1}$ Mpc$^{-1}$ and
$q_{\rm o}=0$.

\section{Comparison to spectral aging methods}
\label{sec:spec}

The age of extragalactic radio sources can be estimated by the use of
spectral aging arguments. This method relies on the determination of
the break in the radio spectrum caused by the time-dependent energy
losses of the relativistic electrons within the cocoon (e.g. Alexander
\& Leahy 1987\nocite{al87}). At frequencies higher than the break
frequency, $\nu_{\rm b}$, the spectrum significantly steepens due to
the radiative energy losses of the electrons. The spectral age of a
electron population, $t_{\rm sa}$, with observed break frequency $\nu
_{\rm b}$ within a magnetic field of strength $B$ which does not vary
in time is proportional to $B^{-3/2} \nu_{\rm b}^{-1/2}$. Despite the
implication of source dynamics that the magnetic field does vary in
time, the constant field relation is usually used in determining the
spectral age of a given source. In general the strength of the
magnetic field will decrease while the source expands and so $t_{\rm
sa}$ will be an overestimate of the true source age,
$t$. However, because of variations of the magnetic field in the
cocoon and the backflow of gas in this region, $t_{\rm sa}$ is usually
found to be lower than the dynamical ages inferred from the advance
speed of the cocoon by various methods (Alexander \& Leahy
1987\nocite{al87}).

The model presented in the following can be viewed as an extension of
the spectral aging formalism. Within the cocoon it traces the
evolution of the local magnetic field in time. This allows the
accurate determination of the energy distribution function of the
relativistic particle population and thus the emitted spectrum at a
given location along the cocoon. It is therefore not surprising that
the model predicts an older age for Cygnus A (see Sect.
\ref{sec:compa}) than the classical spectral index analysis of Carilli
et al. (1991)\nocite{cpdl91}. The combination of a dynamical
model with the accurate treatment of the local evolution within the
cocoon of the radio emission properties implies that the spectral and
dynamical ages are identical.

\subsection{Diffusion of relativistic particles}

In the context of spectral aging methods other processes changing the
energy distribution of the relativistic electrons and thus
invalidating the age estimates have been put forward. These are
summarised and discussed by Blundell \& Rawlings
(2000)\nocite{br00}. They show that most of these processes are rather
inefficient and will not strongly influence spectral aging methods or
the model discussed here. However, Blundell \& Rawlings
(2000)\nocite{br00} claim that the anomalous diffusion mechanism of
Rechester \& Rosenbluth (1978)\nocite{rr78} can lead to very fast
diffusion of relativistic particles through the tangled magnetic field
in the cocoon. So much so that the radio spectra observed at any point
along the cocoon essentially arise from the electrons accelerated by
the jet shock at the hot spot within the last $10^6$ years of the
observation. This would imply that no information on the source age
can be derived from the spatial properties of the observed
emission. Support for a universal particle energy spectrum may come
from the interpretation of observations of Cygnus A by Katz-Stone et
al. (1993)\nocite{kra93}, Rudnick et al. (1994)\nocite{rka94} and
Katz-Stone \& Rudnick (1994)\nocite{kr94}. They find that a single,
non-power law spectrum shifted in frequency by the local strength of
the magnetic field is emitted by all parts of the lobes of Cygnus
A. Note however, that the spectral shape they find may be caused by
the free-free absorption in our own galaxy of the emission of Cygnus A
at low frequencies (Carilli et al. 1991\nocite{cpdl91}. In this case,
the universal spectrum is simply explained by shifting an aging
energy spectrum of relativistic particles which is not a single power
law. Alternatively, changes in the strength of the magnetic field
along a given line of sight may also introduce additional curvature in
the observed spectrum (Rudnick et al. 1994)\nocite{rka94}. The
mentioned interpretation of the radio observations of Cygnus A in
itself is therefore no proof of efficient diffusion acting in the
lobes of radio galaxies.

In the appendix I show that anomalous diffusion is probably much less
effective in the cocoon plasma as previously thought. In general we do
not observe any signature for diffusion losses of the cocoons of radio
sources and so diffusion will not alter the distribution of
relativistic particles within the cocoon. This allows us to use the
spatial distribution of the synchrotron radio emission of FRII sources
to infer their age. The model developed in the following can be viewed
as an extension to the classical spectral aging methods in that it
takes into account the evolution of the magnetic field in the lobe.

\section{The model}
\label{sec:mod}

In this Sect. I briefly summarise the dynamical and radiative model that form
the basis for the extended treatment presented in this paper. Following this
the prescription for the spatial distribution of the synchrotron emission within
the radio lobes is developed.

\subsection{The dynamical model}

\label{sec:dyn}
The large scale structure of radio galaxies and radio-loud quasars of
type FRII is formed by twin jets emanating from the central AGN buried
inside the nucleus of the host galaxy. The jets propagate into
opposite directions from the core of the source. They end in strong
jet shocks and, after passing through these shock, the jet material
inflates the cocoon surrounding the jets. The cocoon is overpressured
with respect to the ambient medium and therefore drives strong bow
shocks into this material.

Falle (1991)\nocite{sf91} and KA showed that the expansion of the bow
shock and the cocoon should be self-similar which is supported by
observations (e.g. Leahy \& Williams 1984\nocite{lw84}, Leahy et
al. 1989\nocite{lms89}, Black 1992\nocite{ab92}). In these models the
density distribution of the material the radio source is expanding
into is approximated by a power law, \( \rho _{\rm{x}}=\rho
_{\rm{o}}(r/a_{\rm{o}})^{-\beta } \), where \( r \) is the radial
distance from the source centre and \( a_{\rm{o}} \) is the core
radius of the density distribution. X-ray observations of groups and
clusters of galaxies show that the density of the hot gas in these
structures is often distributed according to (e.g. Sarazin
1988\nocite{cs88})

\begin{equation}
\label{beta}
\rho _{\rm{x}}=\frac{\rho _{\rm{o}}}{\left[ 1+\left(
r/a_{\rm{o}}\right) ^{2}\right] ^{3\beta '/2}}.
\label{king}
\end{equation}

\noindent Outside a few core radii, $a_{\rm o}$, the power law assumed above with \( \beta =3\beta ' \)
provides an adequate fit to Eq. (\ref{beta}). Even for smaller distances
\( r \) good power law approximations can be found by adjusting \( \beta  \)
and \( \rho _{\rm{o}} \) (e.g. Kaiser \& Alexander 1999a\nocite{ka98a}).

In the model of KA it is also assumed that the rate at which energy is transported
along each jet, \( Q_{\rm{o}} \), is constant and that the jets are in
pressure equilibrium with their own cocoon. The very high sound speed within
the cocoon results in a practically uniform pressure within this region apart
from the tip of the cocoon. The pressure in this `hot spot' region, named for
the very strong radio emission originating in the shock at the end of the jets,
is somewhat higher as the cocoon material injected by the jets at these points
is not yet in pressure equilibrium with the rest of the cocoon.

In the following I will concentrate on only one jet and the half of the cocoon
it is contained in. From KA I take the expressions for the evolution of the
uniform cocoon pressure,

\begin{equation}
\label{pressure}
p_{\rm{c}}=\frac{18c_{1}^{2\left( 5-\beta \right) /3}}{\left( \Gamma _{\rm{x}}+1\right) \left( 5-\beta \right) ^{2}P_{\rm{hc}}}\, \left( \rho _{\rm{o}}a_{\rm{o}}^{\beta }Q_{\rm{o}}^{2}\right) ^{1/3}L_{\rm{j}}^{(-4-\beta )/3},
\end{equation}

\noindent and that of the physical length of the jet,

\begin{equation}
\label{length}
L_{\rm{j}}=c_{1}\left( \frac{Q_{\rm{o}}}{\rho _{\rm{o}}a_{\rm{o}}^{\beta }}\right) ^{1/\left( 5-\beta \right) }t^{3/\left( 5-\beta \right) }.
\end{equation}

\noindent Here, \( \Gamma _{\rm{x}} \) is the ratio of specific heats of
the gas surrounding the radio source, \( t \) is the age of the jet flow and
\( c_{1} \) is a dimensionless constant. The ratio, \( P_{\rm{hc}} \),
of the pressure in the hot spot region and \( p_{\rm{c}} \) is constant
in the model of KA.

KDA extend the model of KA to include the synchrotron emission of the
cocoon. This is done by splitting up the cocoon into small volume
elements, \( \delta V_{\rm{c}} \), the evolution of which is then
followed individually. By assuming that these elements are injected
by the jet into the hot spot region at time \( t_{\rm{i}} \) during a
short time interval \( \delta t_{\rm{i}} \) and then become part of
the cocoon, KDA find
\begin{equation}
\label{element}
\delta V_{\rm{c}}=\frac{\left( \Gamma _{\rm{c}}-1\right) Q_{\rm{o}}}{p_{\rm{c}}\left( t_{\rm{i}}\right) }\left( P_{\rm{hc}}\right) ^{\left( 1-\Gamma _{\rm{c}}\right) /\Gamma _{\rm{c}}}\left( \frac{t}{t_{\rm{i}}}\right) ^{c_{4}}\delta t_{\rm{i}},
\end{equation}

\noindent where \( \Gamma _{\rm{c}} \) is the ratio of specific heats for
the cocoon material and \( c_{4}=\left( 4+\beta \right) /\left[ \Gamma _{\rm{c}}\left( 5-\beta \right) \right]  \).
Because the expansion of the cocoon is self-similar we can set for the volume
of the total volume of the cocoon, following the notation of KA, \( V_{\rm{c}}=c_{3}L_{\rm{j}}^{3} \),
where \( c_{3} \) is a dimensionless constant and depends on the geometric
shape of the cocoon. In order to ensure self-consistency the integration of
Eq. (\ref{element}) over the injection time from \( t_{\rm{i}}=0 \)
to \( t_{\rm{i}}=t \) must be equal to the total cocoon volume \( V_{\rm{c}} \).
Substituting Eq. (\ref{length}) for \( L_{\rm{j}} \) then yields
\begin{equation}
\label{constant}
c_{1}=\left[ P_{\rm{hc}}^{1/\Gamma _{\rm{c}}}\frac{\Gamma
_{\rm{c}}\left( \Gamma _{\rm{c}}-1\right) \left( \Gamma
_{\rm{x}}+1\right) \left( 5-\beta \right) ^{3}}{18c_{3}\left( 9\Gamma
_{\rm{c}}-4-\beta \right) }\right] ^{1/\left( 5-\beta \right) }.
\end{equation}

\noindent Note, that this expression for \( c_{1} \) is different from the
one given by KA. In their analysis they used the expression for conservation
of energy for the entire cocoon

\noindent 
\begin{equation}
\label{energy}
dU_{\rm{c}}=\frac{1}{\Gamma _{\rm{c}}-1}\left[ d\left(
p_{\rm{c}}V_{\rm{c}}\right) +d\left( p_{\rm{h}}V_{\rm{h}}\right)
\right] =Q_{\rm{o}}dt,
\end{equation}

\noindent where \( V_{\rm{h}} \) is the volume of the hot spot region
with pressure \( p_{\rm{h}} \). KA then used the simplifying
assumption that the cocoon has a cylindrical geometry, the expansion
of which is governed along the jet axis by \( p_{\rm{h}} \) while its
growth perpendicular to this direction is driven by \( p_{\rm{c}}
\). This implies $p_{\rm h}/p_{\rm c}=P_{\rm hc}=4R_{\rm T}^2$,
where $R_{\rm T}$ is the ratio of the length of one jet and the full
width of the associated lobe halfway down the jet. Kaiser \&
Alexander (1999b)\nocite{ka98b} subsequently derived empirical fitting
formulae for $P_{\rm hc}$ as functions of $\beta$ and $R_{\rm T}$ from
an analysis of the flow of shocked gas between the bow shock and the
cocoon. Their results showed that the original approximation tends to
overestimate the value of $P_{\rm hc}$. In the following I use a
generalised empirical fitting formula based on their result and
further calculations with additional values of $\beta$,

\begin{equation}
P_{\rm hc}=\left( 2.14 - 0.52 \beta \right) R_{\rm
T}^{2.04-0.25\beta}.
\label{empirical}
\end{equation}

In order to satisfy Eqs. (\ref{constant}) and (\ref{energy}) I
now generalise the approach of KA by setting \( d\left(
p_{\rm{h}}V_{\rm{h}}\right) =fd\left( p_{\rm{c}}V_{\rm{c}}\right) \).
Because of the self-similar expansion of the cocoon \( f \) is a
constant and will depend on the ratio \( P_{\rm{hc}} \). The strict
separation of the cocoon material into the hot spot region and the
rest of the cocoon with two distinctly different values of the
respective pressure within these regions is of course artificial. In
the cocoons of real FRII objects the transition of jet material from
hot spot to cocoon will be accomplished in a continuous hydrodynamical
flow along a pressure gradient much smoother than the sudden change
from \( p_{\rm{h}} \) to \( p_{\rm{c}} \) described here. However, a
detailed model of this flow is beyond the scope of this paper and for
simplicity I will use the assumption of a strict spatial separation in
the following. As the two cocoon regions are in physical contact with
each other, \( f \) may become negative as the expansion of one region
may influence the evolution of the other. Substituting Eqs.
(\ref{pressure}) and (\ref{length}) into Eq. (\ref{energy}) gives
\[
f=\frac{P_{\rm{hc}}^{\left( \Gamma _{\rm{c}}-1\right) /\Gamma _{\rm{c}}}-\Gamma _{\rm{c}}}{\Gamma _{\rm{c}}P_{\rm{hc}}}.\]

\subsection{Synchrotron emission}
\label{sec:syn}

The cocoon volume elements \( \delta V_{\rm{c}} \) are filled with a magnetised
plasma and a population of relativistic electrons accelerated at the shock terminating
the jet flow at the hot spot. They therefore emit synchrotron radio radiation.
{\rm In optically thin conditions the monochromatic luminosity due to this process
can be calculated by folding the emissivity of single electrons with
their energy distribution (e.g. Shu 1991\nocite{fs91}).}

Following KDA I assume that the initial energy distribution of the relativistic
electrons as they leave the acceleration region of the hot spot follows a power
law with exponent \( -p \) between \( \gamma =1 \) and \( \gamma =\gamma _{\rm{max}} \).
The electrons are subject to energy losses due to the adiabatic expansion of
\( \delta V_{\rm{c}} \), the emission of synchrotron radiation and inverse
Compton scattering of the CMB. For a given volume
element at time \( t \) that was injected into the cocoon at time \( t_{\rm{i}} \)
these losses result in an energy distribution (see KDA)
\begin{eqnarray}
\label{distribution}
n'\left( x\right) & = & n_{\rm{o}}\left( t_{\rm{i}}\right) \left(
\frac{2\nu }{3\nu _{\rm{L}}x}\right) ^{-p/2}\left(
\frac{t_{\rm{i}}}{t}\right) ^{c_{4}\left( p+2\right) /3}\nonumber\\
& \times & \left( 1-c_{5}\sqrt{\frac{2\nu }{3\nu _{\rm{L}}x}}\right) ^{p-2},
\end{eqnarray}

\noindent with 
\begin{eqnarray}
\label{loss}
c_{5} & = & \frac{4\sigma _{\rm{T}}}{3m_{\rm{e}}c} t \nonumber\\
& \times & \left\{ \frac{u_{\rm{B}}}{c_{6}}\left[ 1-\left( \frac{t_{\rm{i}}}{t}\right) ^{c_{6}}\right] +\frac{u_{\rm{cmb}}\left( z\right) }{c_{7}}\left[ 1-\left( \frac{t_{\rm{i}}}{t}\right) ^{c_{7}}\right] \right\} ,
\end{eqnarray}

\noindent where \( u_{\rm{cmb}}\left( z\right)  \) is the energy density
of the CMB radiation field at the source redshift \( z \) and \( m_{\rm{e}} \)
is the rest mass of an electron. Here I have used \( c_{6}=1-c_{4}\left( \Gamma _{B}+1/3\right)  \)
and \( c_{7}=1-c_{4}/3 \). The normalisation of the energy spectrum, \( n_{\rm{o}}(t_{\rm{i}}) \),
is given by integrating the initial power law distribution over the entire energy
range
\begin{eqnarray}
\label{norm}
n_{\rm{o}}(t_{\rm{i}}) & = &
\frac{u_{\rm{e}}(t_{\rm{i}})}{m_{\rm{e}}c^{2}}\nonumber\\
&\times&\left[ \frac{1}{2-p}\left( \gamma _{\rm{max}}^{2-p}-1\right) -\frac{1}{1-p}\left( \gamma _{\rm{max}}^{1-p}-1\right) \right] ^{-1},
\end{eqnarray}

\noindent or, for $p=2$,

\begin{equation}
n_{\rm{o}}(t_{\rm{i}})=\frac{u_{\rm{e}}(t_{\rm{i}})}{m_{\rm{e}}c^{2}}\left[
\log \left( \gamma _{\rm max} \right) + \left( \frac{1}{\gamma _{\rm
max}} -1 \right) \right] ^{-1},
\label{normal}
\end{equation}

\noindent where \( u_{\rm{e}} \) is the total energy density of the
relativistic particle distribution. At time \( t_{\rm{i}} \) I set \(
u_{\rm{B}}(t_{\rm{i}})/u_{\rm{e}}(t_{\rm{i}})=r \) and for simplicity
\( p_{\rm{c}}(t_{\rm{i}})=(\Gamma
_{\rm{c}}-1)[u_{\rm{B}}(t_{\rm{i}})+u_{\rm{e}}(t_{\rm{i}})] \).  In
the following I assume that the minimum energy condition (e.g. Miley
1980\nocite{gm80}) is initially fulfilled in each volume element \(
\delta V_{\rm{c}} \) and therefore \( r=(p+1)/4 \). From Eqs.
(\ref{pressure}) and (\ref{length}) it follows that $p_{\rm c} \left(
t_{\rm i} \right) = p_{\rm c} \left( t \right) \left( t / t_{\rm i}
\right)^{c_4 \Gamma_{\rm c}}$ and with the assumption of completely
tangled magnetic fields I find {\rm $u_{\rm B} \left( t \right) =
u_{\rm B} \left( t _{\rm i} \right) \left( t / t_{\rm i} \right)^{c_4
\Gamma _{\rm B}}$} (see also KA). With this the set of equations
describing the radio synchrotron emissivity, $\epsilon
_{\nu}=P_{\nu}/\delta V{\rm c}$, of a given volume element injected
into the cocoon at time $t_{\rm i}$ only depends on the present value
of the pressure in the cocoon, $p_{\rm c}\left( t \right)$, and the
age of the radio source, $t$.

\subsection{Spatial distribution of the emission}
\label{sec:spat}

So far the cocoon volume elements \( \delta V_{\rm{c}} \) were only
characterised by their injection time into the cocoon, \( t_{\rm{i}}
\). From the analysis above it is not possible to decide where they
are located spatially in the cocoon.

From the above analysis it is clear that the radio spectrum emitted by
a given cocoon volume element depends on the `energy loss history' of
this part of the cocoon. Adiabatic losses only change the
normalisation of the emitted spectrum while its slope at a given
frequency is governed by the radiative loss processes of the
relativistic electrons. In the model described above the strength of
the magnetic field which determines the magnitude of synchrotron
losses is tied to the pressure in the cocoon, \( p_{\rm{c}} \). The
value of \( p_{\rm{c}} \) in turn depends on the energy transport rate
of the jet, \( Q_{\rm{o}} \), and a combination of parameters
describing the density distribution of the gas the cocoon is expanding
into, \( \rho _{\rm{o}}a_{\rm{o}}^{\beta } \).  In the analytical
scenario presented here the volume elements are the building blocks of
the cocoon and the variation of the radio surface brightness of the
cocoon of an FRII along its major axis can therefore potentially
provide information on the properties of the source environment. For
this I now identify the cocoon volume elements \( \delta V_{\rm{c}} \)
with infinitesimally thin cylindrical slices with their radius, \( r
\), perpendicular to the jet axis. A similar approach is used in the
model of Chy{\.z}y (1997)\nocite{kc97}. However, in this model the
dynamical evolution and the radio luminosity of the entire cocoon
depend on the geometrical shape and evolution of the volume
elements. In the model of KA and KDA this is not the case and the
identification of the \( \delta V_{\rm{c}} \) with a specific
geometrical shape does not alter their results.

The volume of a thin cylindrical slice of the cocoon is given by \(
\delta V_{\rm{c}}=\pi r_{\rm c}^{2}\delta z \), where \( \delta z \)
is the very small thickness of the slice along the jet axis. The
radius of the slices, \( r_{\rm c} \), depends on their position along
the jet axis, \( l \), which I define in units of \( L_{\rm{j}} \),
the length of one half of the entire cocoon (see Sect.
\ref{sec:dyn}). The outer edges of the slices form the cocoon boundary
or contact discontinuity. Most FRII sources have cocoons of a
relatively undistorted, ellipsoidal shape (e.g. Leahy et
al. 1989\nocite{lms89}). I therefore parameterise the cocoon boundary
as

\begin{equation}
r_{\rm c}=\alpha_{\rm o} \left( 1-l^{\alpha _1} \right)^{\alpha _2},
\label{shape}
\end{equation}

\noindent where $\alpha _{\rm o}$, $\alpha _1$ and $\alpha _2$ can be
determined from radio observations of the cocoon. 

Following the observational results of Leahy \& Williams
(1984)\nocite{lw84} and Leahy et al. (1989)\nocite{lms89} KA and KDA
used the aspect ratio, $R_{\rm T}$, to characterise the geometrical
shape of the cocoons of FRII sources. This ratio is defined as the
length of one side of the cocoon measured from the radio core to the
cocoon tip divided by its width measured half-way along this
line. Using this definition it is straightforward to express $\alpha
_{\rm o}$ in terms of $R_{\rm T}$ as

\begin{equation}
\alpha _{\rm o} = \frac{L_{\rm j}}{2 R_{\rm T}} \left[ 1-\left(
\frac{1}{2} \right) ^{\alpha _1} \right]^{-\alpha _2}.
\label{alpha_exp}
\end{equation}

\noindent The dimensionless volume constant $c_3$ now becomes

\begin{eqnarray}
c_3 & = & \pi \frac{\alpha _{\rm o}^2}{L_{\rm j}^2} \int _0^1 \left( 1-
l^{\alpha _1} \right) ^{2\alpha _2} \, dl \nonumber \\
& = & \pi \frac{\alpha
_{\rm o}^2}{\alpha _1 L_{\rm j}^2} {\rm B}\left( 2 \alpha _2 +1;1/
\alpha _1 \right),
\label{c3_exp}
\end{eqnarray}

\noindent where ${\rm B}(\mu ;\nu)$ is the complete Beta-function.

Tab. \ref{tab:coeff} shows typical values for the dimensionless
constants in the model.

\begin{table}
\begin{center}
\caption{Typical value ranges for the dimensionless constants in the
dynamical model. This assumes $0 \le \beta \le 2$ and $1.3 \le R_{\rm
T} \le 6$ (e.g. Leahy \& Williams 1984).}
\label{tab:coeff}
\begin{tabular}{ccccc}
 $c_1$ & $c_3$ & $c_4$ & $c_6$ & $c_7$\\
\hline\\[-2ex]
$[1.6,6.0]$ & $ [0.01,0.3]$ & $[0.6,1.5]$ & $[-1.5,0.0]$ & $[0.5,0.8]$ \\[0.5ex]
\hline
\end{tabular}
\end{center}
\end{table}

\subsection{Backflow}
\label{sec:back}

The cylindrical slices are injected into the cocoon at a time $t_{\rm
i}$. For simplicity I assume that the slices remain and thus move
within the cocoon as entities afterwards. In other words, I neglect
any mixing of material between slices and I also assume that the
geometrical shape of the slices does not deviate from the initial thin
cylinders. Numerical simulations (e.g. Falle 1994\nocite{sf94}) show
that the gas flow in the cocoon is rather turbulent, at least close to
the hot spot region. It is likely that large-scale turbulent mixing in
the cocoon leads to large distortions of the projected cocoon shape as
seen in radio observations. In this case, the regular cocoon shape
described by Eq. (\ref{shape}) will be a poor representation of
the `true' cocoon shape. For such distorted sources it is unlikely
that the model presented here will provide a good
description. However, the simple picture of cylindrical slices may
still represent the `average' behaviour of the gas flow in more
regularly shaped cocoons rather well.

The model is designed to constrain source and environment parameters
using mainly the gradient of the radio surface brightness along the
cocoon. Problems with this simplified model will therefore arise if
the relativistic electrons in the cocoon are distributed efficiently
by diffusion. In Sect. \ref{sec:disc} I show that the diffusion of
relativistic particles is unlikely to change their distribution on
large scales. It is therefore reasonable to assume that the
relativistic particles are effectively tied to the cocoon slice they
were originally injected into.

Numerical simulations of the large scale structure of FRII sources
strongly suggest that a backflow of material along the jet axis is
established within the cocoon (e.g. Norman et
al. 1982\nocite{nsws82}). The model describing the source dynamics
predicts the growth of the cocoon to be self-similar and therefore the
backflow within the cocoon should be self-similar as well. This
suggests that the position of a slice of cocoon material injected into
the cocoon at time $t_{\rm i}$ is given by $l= \left( t_{\rm i} / t
\right) ^{\alpha _3}$, where $\alpha _3$ governs the velocity of the
backflow at a given position along the cocoon. In order for the model
to be self-consistent, all $\delta V_{\rm c}$ have to add up to the
total volume of the cocoon, $V_c$. Using Eqs. (\ref{pressure}),
(\ref{element}), (\ref{constant}), (\ref{alpha_exp}) and
(\ref{c3_exp}) and replacing $t_{\rm i} / t$ by $l^{1/\alpha _3}\equiv
x$ an implicit Eq. for $\alpha _3$ can be found from this
integration;

\begin{eqnarray}
\lefteqn{\frac{\alpha _1 \Gamma _{\rm c} \left( 5-\beta
\right)}{\left( 9\Gamma _{\rm c} -4-\beta \right)} {\rm B}\left( 2
\alpha _2 +1;1/ \alpha _1 \right) ^{-1}}\nonumber\\ 
& = & \int _0^1
x^{c_4 \left( \Gamma _{\rm c} -1 \right)} \left( 1-x^{\alpha _1 \alpha
_3} \right)^{-2 \alpha _2} dx\\ 
& = & \frac{1}{\alpha _1 \alpha _3}{\rm B} \left( 1-2\alpha _2 ;
\left[ c_4 \left( \Gamma _{\rm c} -1 \right) +1 \right] / \left(
\alpha _1 \alpha _3 \right) \right). \nonumber
\label{alpha3}
\end{eqnarray}

\noindent Note that this expression requires $0 < \alpha _2 < 1/4$ and
gives $\alpha _3 \ge 0$. For the values of the shape parameters
used in Sect. \ref{sec:appl} ($\alpha_1 =2$ and $\alpha_2 =1/3$) and
$\beta =1.5$ I find $\alpha_3 \sim 4.7$. Note that $\alpha _3
\rightarrow \infty$ for $\alpha _2 \rightarrow 1/4$.

The backflow velocity within the cocoon is given by

\begin{equation}
v_{\rm back} = l \left( 1-\frac{5-\beta}{3}\alpha _3 \right) \dot{L}_{\rm
j},
\label{back}
\end{equation}

\noindent where a dot denotes a time derivative. In the rest
frame of the host galaxy the backflow is observed to flow in the
direction of the source core for $v_{\rm back} \le 0$. For $v_{\rm
back} =0$ the cocoon material is stationary in this frame and for
positive values the backflow is strictly speaking not a `backflow' but
trailing after the advancing hot spot. 

The backflow is fastest just behind the hot spot and decelerates along
the cocoon. The deceleration implies a pressure gradient along the
cocoon which may seriously violate the assumption made for the
dynamical model of a constant pressure throughout the cocoon away from
the hot spots. To estimate the magnitude of the pressure gradient I
use Euler's equation

\begin{equation}
\frac{dv_{\rm back}}{dt}=-\frac{1}{\rho _{\rm c}} \frac{dp \left( l
\right)}{L_{\rm j} d l}.
\label{euler}
\end{equation}

\noindent The density of the cocoon material, $\rho _{\rm c}$, is
given by 

\begin{eqnarray}
\rho _{\rm c} & = & \frac{Q_{\rm o} \delta t_{\rm i}}{\left( \gamma _{\rm
j} -1 \right) c^2 \delta V_{\rm c}} \nonumber\\
& = & \frac{p_{\rm c} \left( t_{\rm i}
\right)}{\left( \gamma _{\rm j} -1 \right) \left( \Gamma _{\rm c} -1
\right) c^2} P_{\rm hc}^{\left( \Gamma _{\rm c} -1 \right) / \Gamma
_{\rm c}} l^{c_4/\alpha _3},
\end{eqnarray}

\noindent where I assumed that the entire energy in the jet is
transported in the form of kinetic energy of the flow with a bulk
velocity corresponding to the Lorentz factor $\gamma _{\rm
j}$. With this, it is straightforward with the help of Eq.
(\ref{back}) to integrate Eq. (\ref{euler}) which yields the
pressure along the cocoon as a function of $l$;

\begin{eqnarray}
\frac{p\left( l \right)}{p_{\rm c}} & = & 1-\frac{\left( \alpha' _3 -1
\right) \left( 3 \alpha' _3 + 2 - \beta \right) P_{\rm hc}^{\left(
\Gamma _{\rm c} -1\right) / \Gamma _{\rm c}}}{3 \left( \gamma _{\rm j}
-1 \right) \left( \Gamma _{\rm c} -1 \right) \left[ 2 + c_4/\alpha _3
\left( 1 - \Gamma _{\rm c} \right) \right] } \nonumber \\
& \times & \left( \frac{\dot{L}_{\rm
j}}{c} \right) ^2 l^{2+c_4/\alpha _3 \left(1 - \Gamma _{\rm c} \right)
},
\label{check}
\end{eqnarray}

\noindent where $\alpha' _3 = \left( 5-\beta \right) \alpha _3 / 3$
and I have used the condition $p(0)=p_{\rm c}$. The mean advance speed
of the cocoon of an FRII source, $\dot{L} _{\rm j}$, is inferred from
observations of lobe asymmetries to be in the range from 0.05 c
(Scheuer 1995\nocite{ps95}) to 0.1 c (Arshakian \& Longair
2000\nocite{al00}). This is in good agreement with the predictions of
the dynamical model used here (see KA). The ratio $P_{\rm hc}$ is
usually of order 5 (see Eq. \ref{empirical}) and so even for only
mildly relativistic bulk flow in the jet ($\gamma _{\rm j} \sim 2$)
and large gradients in the backflow velocity, e.g. $\alpha _3 \sim
10$, Eq. (\ref{check}) predicts $0.28 \le p\left( l \right) /
p_{\rm c} \le 1.0$. Note that for $\gamma _j = 5$ the lower limit of
this ratio increases to 0.82.

From this I conclude that the existence of a pressure gradient along
the jet within the cocoon is required to decelerate the backflow of
the cocoon material. However, the pressure varies only by a factor of
a few at most along the entire length of the cocoon. This implies that
within this limit the model is self-consistent.

\section{Comparison with observations}
\label{sec:compa}

The model presented here depends on a large number of parameters. I
assume that each volume element only contains the relativistic
particle population and therefore $\Gamma _{\rm c} = \Gamma _{\rm B} =
4/3$. The other model parameters can be roughly grouped into

\begin{itemize}
\item geometrical parameters: the constants describing the cocoon shape
$\alpha _1$, $\alpha _2$ and $R_{\rm T}$ and the orientation of the
jet axis with respect to the Line-Of-Sight (LOS), $\theta$,
\item properties of the initial energy distribution of the
relativistic electrons and/or positrons: the slope of the
distribution, $p$, and the high energy cut-off, $\gamma _{\rm max}$,
\item properties of the source and its environment: the pressure
within the cocoon, $p_c$, the age of the source, $t$, and the slope of
the power law density distribution of the environment, $\beta$.
\end{itemize}

\noindent In the following I discuss the methods employed in comparing
the model predictions with observations. Using these methods, I will
then show that because of the nature of the model several degeneracies
between model parameters exist. Eliminating these requires further
assumptions to be made but also reduces the complexity of the
parameter estimation.

\subsection{2-dimensional comparison}

A radio map of an FRII source is composed of pixels which contain
information on the monochromatic radio surface brightness of the
source, $S_{\nu} \left( l'_{\rm i}, y'_{\rm j} \right)$, at a given
position $\left( l'_{\rm i}, y'_{\rm j}\right)$ projected onto the
plane of the sky. Here I use the {\em projected} distance along the
jet axis, $l'$, and perpendicular to the jet, $y'$. Both are measured
in units of the projected length of the cocoon, $L'_{\rm j}$. In
optically thin conditions generally present in radio source lobes this
projection corresponds to a LOS integral of the synchrotron
emissivity, $\epsilon _{\nu}$, through the 3-dimensional source at
each pixel location. Following the analysis in Sects. \ref{sec:dyn}
and \ref{sec:syn}, $\epsilon _{\nu}$ is only a function of the
unprojected dimensionless distance from the core measured along the
jet axis, $l$, for a given set of the eight source and environment
parameters. Using the model described above, it is therefore possible
to construct `virtual radio maps' for a given set of model parameters
by projecting the 3-dimensional model onto the plane of the sky. Since
a radio source is in general viewed with its jet axis at an angle
$\theta$ to the LOS, the value of $l$, and therefore that of $\epsilon
_{\nu}$, changes along the path of the LOS integral. Also, the
distance from the core to the tip of the cocoon measured on an
observed radio map, $L'_{\rm j}$, is not equal to the physical size of
the cocoon, $L_{\rm j}=L'_{\rm j}/\sin \theta$. This, of course, also
implies that the aspect ratio of the cocoon, $R'_{\rm T}$, measured in
the observed map is smaller than the `real' value $R_{\rm T} = R'_{\rm
T}/ \sin \theta$, i.e. radio sources appear `fatter' than they really
are. All these projection effects have to be taken into account when
comparing virtual radio maps resulting from these models to
observations.

The surface brightness is calculated at a frequency $\nu$ which is
given by the observing frequency in the frame of the observer, $\nu'$,
and the redshift of the source, i.e. $\nu=(1+z) \nu'$. Finally, to
account for cosmological effects the result of the LOS integration
must be multiplied by $(1+z)/D_{\rm L}^2$, where $D_{\rm L}$ is the
luminosity distance to the source.

The model map can be calculated at arbitrary resolution. However,
before comparing the result with the observed map it must be convolved
with the beam of the radio telescope used for the observations. This
was done assuming a 2-dimensional Gaussian shape for the telescope
beam. 

Once a virtual map is compiled for a set of model parameters this map
can be compared pixel by pixel to a map resulting from observations of
an FRII radio source using a $\chi ^2$-technique, 

\begin{equation}
\chi _{\nu}^2=\sum\limits^n_i \sum\limits^m_j \frac{
\left[S_{\nu}\left(l'_{\rm i}, y'_{\rm j} \right)-M_{\nu}\left(l'_{\rm
i}, y'_{\rm j} \right) \right] ^2}{\sigma _{\nu}^2}.
\end{equation}

\noindent Here, $S_{\nu}$ is the measured monochromatic surface
brightness with rms error $\sigma _{\nu}$ and $M_{\nu}$ is the model
prediction. 

It is then possible to find the best-fitting model by varying the
model parameters and thereby minimising the resulting $\chi
^2$-difference between virtual and observed map. The minimisation
routine uses a n-dimensional downhill simplex method (Press et
al. 1992\nocite{ptvf92}), where n is the number of model parameters to
be fitted. The minimisation can be done separately for the two halves
of each cocoon since the model describes one jet and the associated
half of the cocoon. Although in principle the minimisation can be done
with one observed map at a single observing frequency, the constraints
on the model parameters are improved by using two maps at two
different frequencies. In this case the $\chi ^2$-differences
resulting from the two maps are simply added together. In
principle this is equivalent to compiling spectral index maps from two
observed maps and comparing these with the model predictions. However,
using the two maps directly has the advantage that pixels below the
rms limit in one map but not in the other are not entirely lost for
the fitting procedure. Furthermore, information on the absolute
surface brightness in a given location is not contained in a spectral
index map. The model would have to be normalised `by hand' and the
various possibilities to do this would lead to ambiguities in the
estimation of model parameters. In Sec. \ref{sec:appl} I use two
individual maps at two frequency for each source.

Using the best-fitting model parameters $p_{\rm c}$ and $t$, the
energy transport rate of the jet, $Q_{\rm o}$, and the parameter
combination $\rho _{\rm o} a_{\rm o}^{\beta}$ describing the density
distribution in the source environment can be calculated from Eqs.
(\ref{pressure}) and (\ref{length}).

Preliminary results using this comparison technique applied to radio
observations of Cygnus A was presented in Kaiser
(2000)\nocite{ck00}. In Sect. \ref{sec:appl} this analysis is
extended to also include 3C 219 and 3C 215.

\subsection{1-dimensional comparison}

The 2-dimensional comparison method described in the previous section
requires a ray-tracing algorithm for the projection of the
3-dimensional model. For many maps of radio sources the number of
pixels are so large that this method can become computationally very
expensive. Furthermore, for maps of lower resolution the cocoon may
not be resolved in the direction perpendicular to the jet axis. In
such maps the surface brightness gradient along the jet axis can be
extracted by taking a cut through the map along the line connecting
the source core and the radio hot spot in the cocoon on one side. This
yields a 1-dimensional curve of surface brightness as a function of
projected distance from the source core, $S_{\nu} \left( l'
\right)$. The off-axis pixels will not add much information. The
function $S_{\nu} \left( l' \right)$ can then be compared with the
model predictions and a best-fitting model may be found using two maps
at two observing frequencies as outlined in the 2-dimensional
case. This 1-dimensional comparison involves a much smaller number of
pixels for which a model prediction must be calculated than the
2-dimensional method.

\subsection{Model parameters}
\label{sec:degen}

\begin{table}
\begin{center}
\caption{Model parameters of the fiducial model (see text).}
\label{tab:modpara}
\begin{tabular}{lcc}
\hline\\[-2ex]
& $p_{\rm c} / {\rm J \ m}^{-3}$ & $5\cdot10^{-11}$\\
& $t / {\rm years}$ & $10^7$\\
& $\theta / {\rm degrees}$ & $60$\\
& $R_{\rm T}$ & $2.3$\\
& $\alpha _1$ & $2$\\
\raisebox{1.25ex}[-2.25ex]{fixed parameters}& $\alpha _2$ & $1/3$\\
& $p$ & $2.0$\\
& $\gamma _{\rm max}$ & $10^{4.5}$\\
& $\beta$ & $1.5$\\
& $L_{\rm j} / {\rm kpc}$ & $100$\\[0.5ex]
\hline\\[-2ex]
& $B / \mu {\rm G}$ & $130$\\
& $Q_{\rm o} / {\rm W}$ & $4.3 \cdot 10^{38}$\\
derived parameters & $\rho _{\rm o} a_{\rm o}^{\beta} / {\rm kg \ m}^{-1.5}$ & $8.8 \cdot
10^7$\\
& $L'_{\rm j} / {\rm kpc}$ & $86.6$\\
& $R'_{\rm T}$ & $2$\\
& $\alpha _3$ & $4.7$\\[0.5ex]
\hline
\end{tabular}
\end{center}
\end{table}

To study the influence of the individual model parameters on the
surface brightness distribution, I define a fiducial model with a set
of fixed model parameters given in Tab. \ref{tab:modpara}. These model
parameters imply a magnetic field just behind the hot spot of 13 nT or
130 $\mu$G. The highest frequency of the emitted synchrotron spectrum
is then $\sim 300$ GHz which is close to the usually assumed high
frequency cut-off in the spectrum of radio galaxies in minimum energy
arguments (e.g. Miley 1980\nocite{gm80}). The density parameter $\rho
_{\rm o} a_{\rm o}^{\beta}=8.8 \cdot 10^7$ kg m$^{-1.5}$ which
corresponds to a central density of $5 \cdot 10^{-22}$ kg m$^{-3}$ or
0.3 particles per cm$^{-3}$ if $a_{\rm o}=10$ kpc. I also assume that
the jet of the sources is 100 kpc long and is viewed at an angle of
$\theta =60^{\circ}$ to the line of sight. This implies that it would
be observed to have a length of $L'_{\rm j} =86.6$ kpc corresponding
to 31.2" at a redshift $z=0.1$. For a measured $R'_{\rm T}=2$ the
aspect ratio of the cocoon is $R_{\rm T}=2.3$ for the assumed viewing
angle.

For simplicity and ease of comparison I use the 1-dimensional model
which only predicts the surface brightness distribution along the jet
axis (see previous Sect.). The variation of the model predictions
with varying parameters for the 2-dimensional case are essentially
similar but the differences between maps are more difficult to
visualise.

\begin{figure}
\centerline{
\epsfig{file=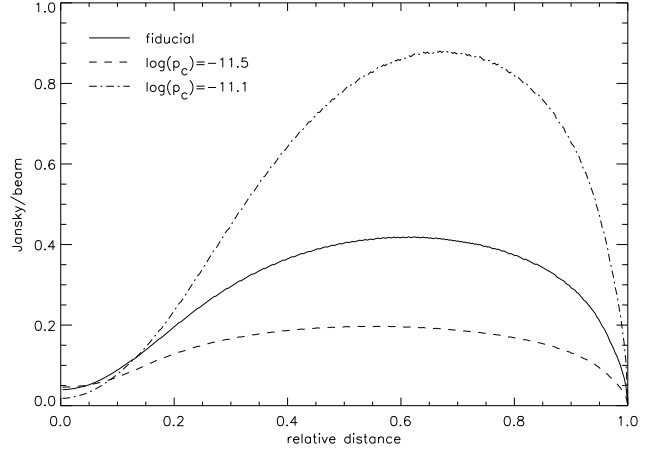, width=9cm}}
\caption{The influence of the cocoon pressure on the radio surface
brightness along the jet axis. All curves are plotted for the
parameters of the fiducial model except for a variation in the cocoon
pressure. The relative distance from the core of the source is given
in units of $L'_{\rm j}=86.6$ kpc (see text). The surface brightness
is plotted as it would be extracted by an observer from a map of a
source at redshift $z=0.1$.}
\label{fig:pccomp}
\end{figure}

The solid line in Fig. \ref{fig:pccomp} shows $S_{\nu} \left( l
\right)$ at $\nu'=1.5$ GHz assuming a pixel size of 0.3"$\times$0.3",
corresponding to 0.8 kpc$\times$0.8 kpc for $z=0.1$, appropriate for a
telescope beam of 1.2" FWHM. For simplicity and in contrast to the
2-dimensional comparison method the averaging effects of the observing
beam which extends over four pixels on the curve was not taken into
account. In any case, in this Sect. the only interest is in gross
trends of the model predictions when the model parameters are varied
and the effective smoothing of the beam on the already rather smooth
curve is small. A continuous curve is shown, since the pixel size is
small compared to the scale of the plot. The length of the cocoon
corresponds to more than 100 pixels. As pointed out in Sect.
\ref{sec:syn}, the emission of the hot spot is not modeled in the
approach presented here. The cocoon surface brightness shown in this
and the following figures is caused by the cocoon material only after
it has passed through the hot spot region. The emission from the hot
spot will in general dominate the total emission from the end of the
cocoon at $l\rightarrow 1$. In a comparison with observed maps the
predictions of the model can therefore not be used in this region.

Also shown in Fig. \ref{fig:pccomp} are the results for the same
model with a higher and lower cocoon pressure. In the model the
strength of the magnetic field and the energy density of the
relativistic particles in the cocoon is tied to the cocoon pressure. A
higher pressure therefore causes a higher peak of the surface
brightness distribution. However, the increased synchrotron energy
losses of the relativistic electrons also lead to a stronger gradient
of the distribution towards the core of the source, i.e. the older
parts of the cocoon.

\begin{figure}
\centerline{
\epsfig{file=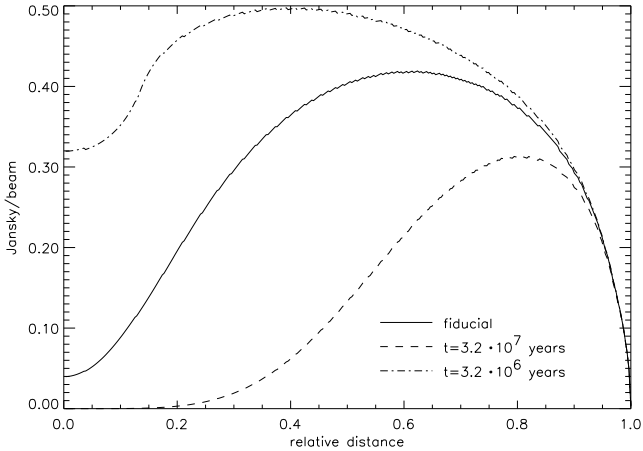, width=9cm}}
\caption{The influence of the source age on the radio surface
brightness. All model parameters as in Fig. \ref{fig:pccomp} except
for a variation in the source age.}
\label{fig:tcomp}
\end{figure}

In Fig. \ref{fig:tcomp} the effects of the source age on the
distribution of $S_{\nu} \left( l \right)$ is shown. The population of
relativistic particles at a given relative distance, $l$, from the
core of the source has spent more time in the cocoon in an old source
compared to a young source. This implies stronger energy losses and
the peak of the surface brightness distribution is therefore located
closer to the hot spot in an old source. Furthermore, in an
observation with a given detectable threshold of $S_{\nu} \left( l
\right)$ a larger fraction of the cocoon of a young source will be
visible than of an older source. 

\begin{figure}
\centerline{
\epsfig{file=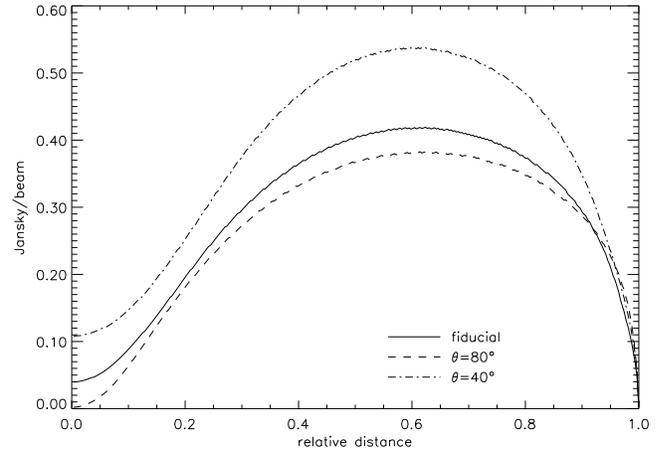, width=9cm}}
\caption{The influence of the viewing angle on the radio surface
brightness. All model parameters as in Fig. \ref{fig:pccomp} except
for a variation in $\theta$. Note that the measured length of the
cocoon, $L'_{\rm j}$, changes from 86.6 kpc for the fiducial model to
98.5 kpc for $\theta = 80^{\circ}$ and 64.3 kpc for $\theta =
40^{\circ}$.}
\label{fig:avcomp}
\end{figure}

The same source viewed at different angles of the jet axis to the LOS,
as shown in Fig. \ref{fig:avcomp}, results in a scaling of the surface
brightness similar to the effects of changing the cocoon
pressure. Note however, that the overall normalisation of $S_{\nu}
\left( l \right)$ depends sensitively on the value of the cocoon
pressure. The effects of a higher pressure also include a steepening
of $S_{\nu} \left( l \right)$ starting from the peak of this function
towards the source core (Fig. \ref{fig:pccomp}). This is not seen for
variations of the viewing angle (Fig. \ref{fig:avcomp}). In practice
therefore, the best-fitting values for the cocoon pressure is set
mostly by the overall surface brightness of the cocoon while the
viewing angle is mainly determined by the behaviour of $S_{\nu} \left(
l \right)$ close to the core.

\begin{figure}
\centerline{
\epsfig{file=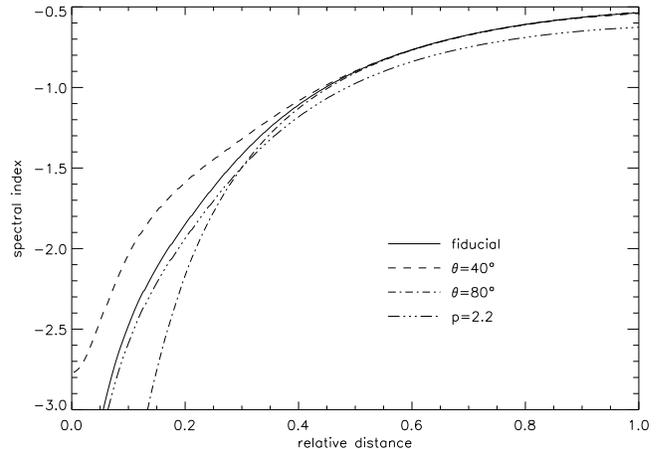, width=9cm}}
\caption{Spectral index between 1.5 GHz and 5 GHz for the fiducial
model and models with varying viewing angle, $\theta$, and slope of
the initial energy spectrum of the relativistic particles, $p$. All
other model parameters as in Fig. \ref{fig:pccomp}.}
\label{fig:speccomp}
\end{figure}

\begin{figure}
\centerline{
\epsfig{file=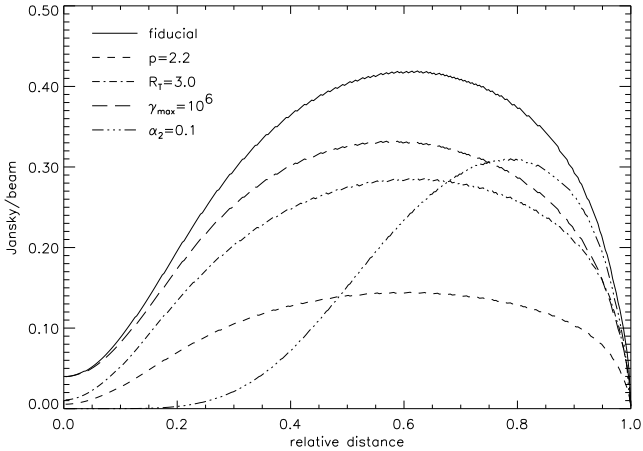, width=9cm}}
\caption{Effects of other model parameters on the radio surface
brightness. Model parameters are as in Fig. \ref{fig:pccomp} unless
otherwise indicated in the legend.}
\label{fig:multicomp}
\end{figure}

Fig. \ref{fig:multicomp} summarises the effects of the remaining
model parameters on the surface brightness distribution. Changing the
aspect ratio of the cocoon, $R_{\rm T}$, results in a scaling of
$S_{\nu} \left( l \right)$ very similar to the effects of a changing
viewing angle, $\theta$. Decreasing $\alpha _2$ leads to the end of
the cocoon to become more blunt which in turn implies a lower value of
$\alpha _3$, i.e. a slower backflow within the cocoon (Eqs.
\ref{alpha3} and \ref{back}). This is analogous to an older particle
population at a given distance from the hot spot and the resulting
surface brightness distribution is similar to that of an older
source. Changing the second shape parameter $\alpha _1$ has similar
effects as that of a variation of $\alpha _2$ and is therefore not
shown.

The main effects of changing the slope, $p$, and the high energy
cut-off of the initial energy distribution of the relativistic
electrons, $\gamma _{\rm max}$, is a change in the scaling of $S_{\nu}
\left( l \right)$ (see Fig. \ref{fig:multicomp}). This is caused by
the dependence of the normalisation of the energy spectrum on $p$ and
$\gamma _{\rm max}$, Eq. (\ref{normal}). Both scaling effects are
similar to the effects of a variation of the viewing angle,
$\theta$. The effect of a steeper energy spectrum also causes an
off-set in the distribution of the spectral index as a function of $l$
as shown in Fig. \ref{fig:speccomp}. However, in the cocoon region
closest to the source centre and least influenced by the hot spot
emission, this effect is small compared to that of a variation of the
viewing angle, $\theta$.

Finally, a different value of the slope of the gas density
distribution in the source environment, $\beta$, has a negligible
effect on $S_{\nu} \left( l \right)$. The relevant curves for $\beta
=1$ and $\beta =2$ are indistinguishable from that for $\beta =1.5$
shown in Figs. \ref{fig:pccomp} to \ref{fig:multicomp}.

\subsection{Degeneracy of parameters}

Both, $\alpha _1$ and $\alpha _2$, control how pointed the shape of
the cocoon is. The ends of the cocoon are dominated by the emission
from the hot spots which are not part of the model. Furthermore, if
the jet direction is not stable over the lifetime of the source, they
may at times advance ahead of the rest of the cocoon which distorts
the cocoon shape. This has been referred to as the `dentist drill'
effect (Scheuer 1982\nocite{ps82}). In the previous Sect. (see also
Eq. \ref{alpha3}) it was already shown that changing $\alpha _1$
and/or $\alpha _2$ results in a change of the profile and magnitude of
the backflow in the cocoon. This in turn causes changes to the surface
brightness profiles similar to changes of the source age. For these
reasons the model cannot provide strong constraints on either of these
parameters, at least not independent of the source age. Therefore I
set in the following $\alpha _1 =2$ and $\alpha _2 =1/3$.

The surface brightness distribution predicted by the model is almost
independent of the slope of the density profile of the external
medium, $\beta$. I therefore set $\beta =1.5$ without influencing the
model results significantly. Note that because of Eq.
(\ref{alpha3}) this then also implies a fixed value for $\alpha _3
=4.8$. This may seem a rather significant restriction of the model but
as was shown in the previous Sect. $\alpha _1$ or $\alpha _2$,
through their influence on $\alpha _3$, and the age of the source,
$t$, are degenerate model parameters. In the absence of geometrical
constraints on $\alpha _1$ or $\alpha _2$, setting them to reasonable
values allows the determination of the source age from the fitting
method. The ages derived from the model will therefore always depend
somewhat on the choice for the source geometry.

The slope of the initial energy distribution of the relativistic
particles, $p$, mainly influences the spectral index distribution in
the cocoon. Observations suggest that $2\le p \le3$ and the model is
therefore restricted to values in this range. It follows that the
model is rather insensitive to the exact value of the high energy
cut-off of the distribution, $\gamma _{\rm max}$. A value of $\gamma
_{\rm max}$ much smaller than the $10^{4.5}$ used in the fiducial
model will, in addition to a change of the overall scaling, cause the
emission region to shorten along the jet axis at a given observing
frequency. However, because the highest frequency of the synchrotron
radiation of the cocoon, $\nu_{\rm max}$, depends on $\gamma _{\rm
max}^2$ a small value of the high energy cut-off also implies a
significantly smaller $\nu_{\rm max}$. Values substantially below
$\nu_{\rm max}\sim 100$ GHz, which is assumed, here are unlikely in
view of observations. I therefore set $\gamma _{\rm max} = 10^{4.5}$.

As was pointed out above, the effects of varying the initial slope of
the energy distribution of the relativistic particles, $p$, are small
in general. In Sect. \ref{sec:appl} the 1 and 2-dimensional
comparison methods are applied to observations of three FRII-type
radio sources. It is shown there that in almost all cases the
best-fitting model parameters for the 2-dimensional method require $p$
to be close to 2. To prevent the degeneracy between $p$ and $\theta$
to influence the model fits in the 1-dimensional method which involves
fewer degrees of freedom, I set $p=2$ in this case.

Finally, the degeneracy between $R_{\rm T}$, the width of the cocoon,
and the viewing angle, $\theta$, is resolved by determining the
projected cocoon width $R'_{\rm T}$ from the observed map and use
$R_{\rm T} = R'_{\rm T}/\sin \theta$ in the model calculations.

Using these additional assumptions, the number of model parameters
which are fitted by comparison to the observations decreases to four:
The pressure in the cocoon, $p_{\rm c}$, the source age, $t$, the
viewing angle, $\theta$, and the initial slope of the energy
distribution of the relativistic particles, $p$. In the case of the
1-dimensional comparison method $p$ is fixed to a value of 2.

\section{Application to Cygnus A, 3C 219 and 3C 215}
\label{sec:appl}

To test the model predictions for the source environment against
direct X-ray observations over a range of different viewing angles,
the radio data of three different FRII-type objects are used: the
narrow-line radio galaxy Cygnus A, the broad-line radio galaxy 3C 219
and the radio-loud quasar 3C 215. According to orientation-based
unification schemes of the various sub-classes of radio-loud AGN
(e.g. Barthel 1989\nocite{pb89}), the viewing angle, $\theta$, of
Cygnus A should be greater than those of 3C 219 and 3C
215. Furthermore, 3C 219 and 3C 215 were selected because of their
rather irregular radio lobe structure. The model is based on a very
regular geometrical shape of the cocoon, Eq. (\ref{shape}), and
using 3C 219 and 3C 215 it is possible to estimate to what extent this
restriction limits the applicability of the model.

Note, that results of the model fitting for Cygnus A are
presented in \ref{fig:least} and \ref{fig:heast} only at one
frequency. However, the model fits are always obtained for all sources
using two maps at two different frequencies.

\subsection{Cygnus A}

\begin{figure}
\centerline{
\epsfig{file=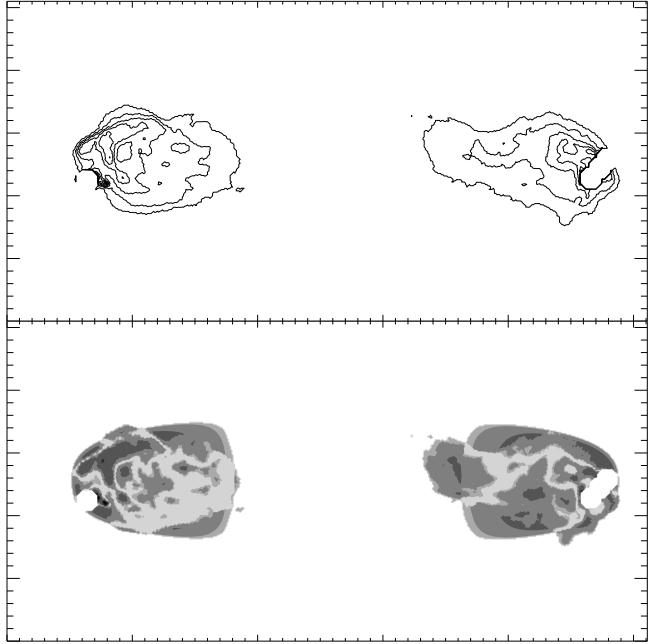, width=8.5cm}}
\caption{The 2-dimensional comparison method applied to Cygnus A. The
upper panels shows the observed map at 1.8 GHz. Contours are
increasing linearly in steps of 0.43 Jy beam$^{-1}$, starting at 0.14
Jy beam$^{-1}$, the 5-$\sigma$ noise level, to a peak just above 3 Jy
beam$^{-1}$. The map is rotated by about 20$^{\circ}$ clockwise
compared to the true position angle. The lower panel shows the
$\chi^2$-deviation of the best-fitting models. The filled contours
show regions where $\chi^2=0$ (white), $0< \chi^2 \le 1$, $1 < \chi^2
\le 2$, $2 < \chi^2 \le 10$, $10 < \chi^2 \le 50$, $50 < \chi^2 \le
100$ and $\chi^2 > 100$ (black). Note that the best-fitting model for
the eastern and western lobes are not identical (see Table
\ref{tab:res}). The results for both lobes are presented together for
ease of comparison. The best-fitting model is obtained by fitting
the 1.8 GHz map presented here and simultaneously the 5 GHz data (see
Fig. \ref{fig:compac})}.
\label{fig:compa}
\end{figure}

\begin{figure}
\centerline{
\epsfig{file=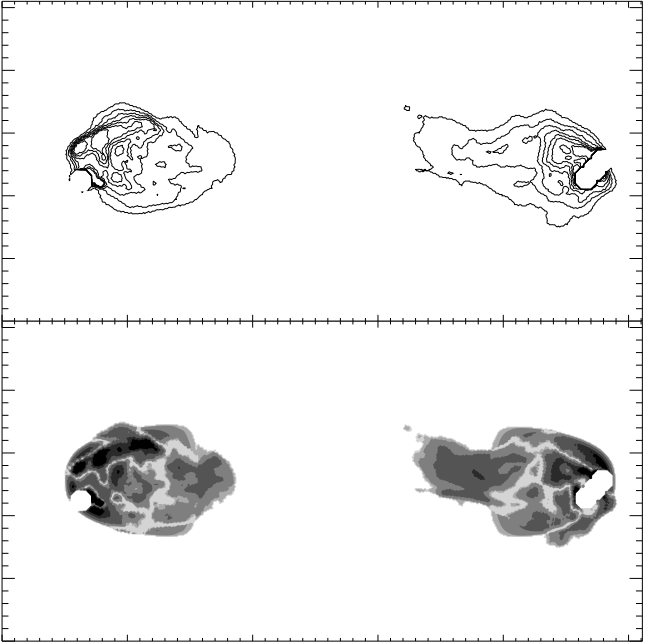, width=8.5cm}}
\caption{Same as Fig. \ref{fig:compa} but at 5 GHz. The contour levels
in the upper panel are spaced linearly in steps of 0.12 Jy beam$^{-1}$
starting at 0.03 Jy, the $5-\sigma$ level, to 0.76 Jy
beam$^{-1}$. Higher contours are omitted for clarity. The $\chi
^2$-contours are as in Fig. \ref{fig:compa}.}
\label{fig:compac}
\end{figure}

For this source at $z=0.056$ I used raw data from the VLA archive at
1.8 GHz and 5 GHz. The lower frequency observations were made in
August/September 1987 in A-array and the 5 GHz data were obtained in
January 1984 in B-array. A detailed analysis of these and other
observations of Cygnus A can be found in Carilli et
al. (1991)\nocite{cpdl91}. Standard calibration and self-calibration
was performed using the software package AIPS. This resulted in two
radio maps with comparable angular resolution of 1.3" at the two
observing frequencies. The CLEAN components were restored in maps with
an individual pixel size of 0.3"$\times$0.3". The map at 1.8 GHz with
the hot spot emission removed (see below) is shown in the upper panel
of Fig. \ref{fig:compa}. The rms noise in the maps is 0.03 Jy
beam$^{-1}$ at 1.8 GHz and 0.006 Jy beam$^{-1}$ at 5 GHz. All pixels
in the maps with a surface brightness below 5$\sigma$ were discarded
for the 2-dimensional comparison method. In the case of the
1-dimensional method only pixels below 3-$\sigma$ were neglected. The
surface brightness distribution along the jet axis, $S_{\nu} \left( l
\right)$, for the 1-dimensional method was obtained for both radio
lobes along a cut from the core of the source to the eastern and
western hot spot respectively. 

Since the hot spot emission is not modeled, it has to be removed from
the maps and the 1-dimensional cuts. In the maps an aperture centered
on the surface brightness peak in each lobe and with a radius of 2.6"
corresponding to twice the beam width is removed (see Fig.
\ref{fig:compa}). In the western lobe the bright secondary hot spot
(Carilli et al. 1991\nocite{cpdl91}) and the bright ridge connecting
this hot spot with the main one are also removed. For the
1-dimensional method, the distance of the hot spot to the edge of the
lobe, $\Delta l$, was estimated as the distance of the maximum of
$S_{\nu} \left( l \right)$ to the last point at which this function
has a value above 3$\sigma$ in the direction away from the source
core. To remove the contribution of the hot spots to $S_{\nu} \left( l
\right)$, all pixels within $2\Delta l$ of the edge of the lobe were
neglected in the following 1-dimensional comparison process.

The spatial resolution of the radio maps of Cygnus A is comparatively
high. The 1.3" angular resolution corresponds to a spatial resolution
of about 1.9 kpc. For many sources, particularly at high redshift,
maps of such high quality are not available. In order to estimate the
effects of a lower angular and therefore lower spatial resolution, I
also convolved the two maps of Cygnus A with a Gaussian beam of 5"
FWHM. In the case of the 2-dimensional method the radius of the
aperture used to remove the hot spot emission was fixed to 5". For the
1-dimensional method, the surface brightness distribution along the
jet axis was extracted from these lower resolution maps in the same
way as in the higher resolution case. Note that the emission of the
hot spots is smeared out over a larger area in the low resolution
maps. To avoid any bias from the enhanced emission of the hot spot
region I used the higher value of $2\Delta l$ obtained from the low
resolution maps in both, the high and low resolution, 1-dimensional
comparison.

\begin{table*}
\begin{center}
\caption{The best-fitting model parameters from the 1 and
2-dimensional comparison methods.}
\label{tab:res}
\begin{tabular}{cllcccc}
& & & $\log \left( p_{\rm c} / {\rm J m^{-3}} \right)$ &
$t/10^6{\rm years}$ & $\theta /{\rm degrees}$ & $p$\\
\hline\\
& & 2-D & $-11.01^{+.06}_{-.01}$ & $14.5^{+2.9}_{-1.3}$ &
$71^{+16}_{-29}$ & $2.00^{+.07}_{-.00}$\\[1.5ex]
& \raisebox{2.25ex}[-2.25ex]{5" resolution} & 1-D & $-11.08^{+.04}_{-.02}$ &
$13.8^{+4.0}_{-0.9}$ & $79^{+10}_{-18}$ & \ldots\\[1.5ex]
\raisebox{2.25ex}[-2.25ex]{Cygnus A, eastern lobe} & & 2-D & $-10.90^{+.02}_{-.05}$ & $13.8^{+2.5}_{-.9}$ &
$87^{+3}_{-11}$ & $2.07^{+.01}_{-.07}$\\[1.5ex]
& \raisebox{2.25ex}[-2.25ex]{1.3" resolution} & 1-D & $-11.07^{+.05}_{-.08}$ & $14.5^{+5.5}_{-1.3}$ &
$84^{+6}_{-28}$ & \ldots\\[3ex]
& & 2-D & $-10.85^{+.04}_{-.25}$ & $15.2^{+16.5}_{-1.7}$ &
$75^{+14}_{-56}$ & $2.13^{+.04}_{-.13}$\\[1.5ex]
& \raisebox{2.25ex}[-2.25ex]{5" resolution} & 1-D & $-11.01^{+.05}_{-.02}$ &
$15.9^{+7.1}_{-2.4}$ & $83^{+7}_{-21}$ & \ldots\\[1.5ex]
\raisebox{2.25ex}[-2.25ex]{Cygnus A, western lobe} & & 2-D & $-10.87^{+.05}_{-.11}$ & $17.4^{+3.6}_{-2.6}$ &
$82^{+2}_{-4}$ & $2.10^{+.06}_{-.1}$\\[1.5ex]
& \raisebox{2.25ex}[-2.25ex]{1.3" resolution} & 1-D & $-11.09^{+.08}_{-.13}$ & $15.2^{+6.7}_{-2.6}$ &
$79^{+11}_{-25}$ & \ldots\\[3ex]
& & 2-D & $-12.12^{+.09}_{-.04}$ & $34.8^{+.8}_{-7.8}$ & $59^{+12}_{-6}$ & $2.03^{+.09}_{-.01}$\\[1.5ex]
\raisebox{2.25ex}[-2.25ex]{3C 219} & \raisebox{2.25ex}[-2.25ex]{4.3" resolution} & 1-D & $-11.91^{+.03}_{-.08}$ & $44.8^{+17.0}_{-9.2}$ &
$65^{+24}_{-14}$ & \ldots\\[3ex]
& & 2-D & $-12.20^{+.39}_{-.37}$ & $35.6^{+23.4}_{-22.1}$ &
$31^{+32}_{-21}$ & $2.22^{+.28}_{-.22}$\\[1.5ex]
\raisebox{2.25ex}[-2.25ex]{3C 215} & \raisebox{2.25ex}[-2.25ex]{1.9" resolution} & 1-D & $-12.38^{+.08}_{-.06}$ & $38.1^{+9.9}_{-20.7}$ & $43^{+57}_{-5}$ & \ldots\\[0.5ex]
\hline
\end{tabular}
\end{center}
\end{table*}

\subsubsection{The eastern lobe}

\begin{figure}
\centerline{
\epsfig{file=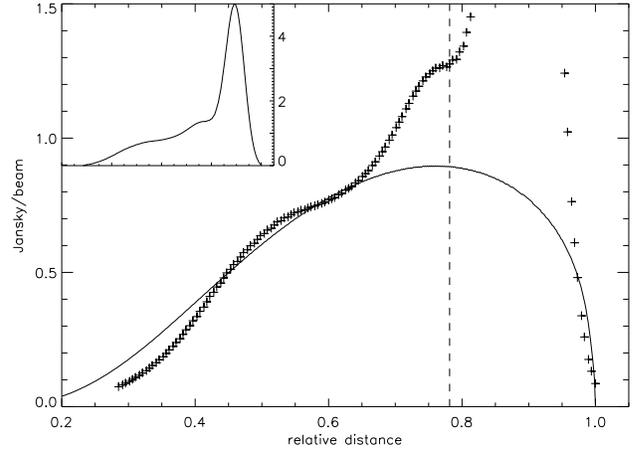, width=9cm}}
\caption{Comparison of the predicted surface brightness distribution
along the cocoon with the low resolution observations of the eastern
lobe of Cygnus A. The crosses show the value of $S_{\nu} \left( l
\right)$ at 1.8 GHz along the jet axis taken from the VLA map
convolved with a 5" Gaussian beam. The solid line shows the
best-fitting model with the model parameters given in table
\ref{tab:res}. Only data points left of the dashed line located at
$2\Delta l$ from the tip of the cocoon (see text) were used in the
fitting procedure to avoid the contribution of the hot spot. The
entire observed range of $S_{\nu} \left( l \right)$ including the peak
of the hot spot is shown in the inset.}
\label{fig:least}
\end{figure}

\begin{figure}
\centerline{
\epsfig{file=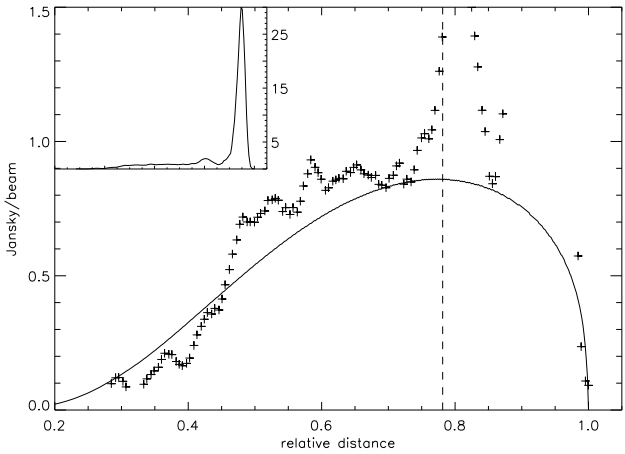, width=9cm}}
\caption{Comparison of the predicted surface brightness distribution
along the cocoon with the high resolution observations of the eastern
lobe of Cygnus A. The data is taken from the unconvolved VLA map. As
in Fig. \ref{fig:least} the solid line shows the best-fitting model
and only data left of the dashed line was used in the fitting
procedure.}
\label{fig:heast}
\end{figure}

The eastern lobe of Cygnus A is covered by 46.6 independent telescope
beams along the jet while in the widest part there are 22.2 beams
across. In the lower resolution maps these numbers decrease to 12.9
and 6.1, respectively. Note, that only a fraction of the lobes has an
observed surface brightness above the rms limits, i.e. they do not
extend all the way from the hot spots to the core in the
observations. This implies that the model fits are based on regions
covering less area than the theoretical extend of the cocoons. For
both resolutions I find an axial ratio, $R'_{\rm T}$, of 2.1 at the
point where the lobe is widest from the 2-dimensional maps. The length
of the lobe, $L'_{\rm j}$, is 64.4" for the low resolution map and
60.6" for the high resolution map. The prediction of the
best-fitting model in comparison with the observations is shown in the
lower panels of Figs. \ref{fig:compa} and \ref{fig:compac} for the
2-dimensional method and in Figs. \ref{fig:least} and \ref{fig:heast}
for the 1-dimensional method at 1.8 GHz. The 5 GHz data is not shown
for the 1-dimensional method but is similar to the result at 1.8
GHz. The parameters of the best-fitting models are given in table
\ref{tab:res}. The errors on these and for all the following model
fits are estimated using the boot-strap method (e.g. Press et
al. 1992\nocite{ptvf92}). It is not possible to calculate error
estimates using the $\chi ^2$-values derived in the minimisation
procedure directly as the values of the surface brightness in
neighboring pixels are not independent. Roughly 2000 data sets were
created by drawing data with replacement from the original set. The
same minimisation procedure as in the original model fitting was then
applied to them and the error given in the table is the 1$\sigma$
limit on the respective model parameters.

Even after subtracting the contribution of the hot spot the deviations
of the model from the observations are large. The fact that the model
fit is poorer at 5 GHz is mainly caused by the smaller rms noise of
the observed 5 GHz map. As expected, the model cannot fit structures
which appear as discrete surface brightness enhancements in the
maps. This is particularly clear in the case of the bright arc seen
just behind the hot spot in radio maps of the eastern lobe (see
Figs. \ref{fig:compa} and \ref{fig:compac}) which also causes the
secondary peak in $S_{\nu} \left( l \right)$ at about $l=0.83$ (see
Fig. \ref{fig:heast}). Although convolving the maps with a larger beam
`draws' some flux from the hot spot into the arc mentioned above, the
results for the two different resolutions are very similar for both
comparison methods. At both frequencies the observed maps show a
concentration of the radio emission towards the centres of both
lobes. This region is also extended a long way along the jet axis,
particularly in the western lobe (see following section). In
Figs. \ref{fig:compa} and \ref{fig:compac} it is clear that the model
cannot fit this concentration properly. The very smooth lobes of the
model are `fatter' than the observed lobes further away from the hot
spots and do not extend as far back as the observations indicate. This
is particularly striking in the western lobe at 5 GHz (see
Fig. \ref{fig:compac}). This clearly illustrates the limitations of
the model assumption of a smooth shape of the cocoon and a regular
backflow within the cocoon. The large uncertainty of the viewing
angle, $\theta$, is caused by the model mainly depending on $\sin
\theta$ which changes only by a factor 1.2 within the estimated
errors. This is comparable to the uncertainties of the other model
parameters.

\begin{table*}
\begin{center}
\caption{Properties of the source environment derived from the
best-fitting model parameters in comparison with observations. For the
determination of $\rho _{\rm o}$ the core radii of Hardcastle \&
Worrall (2000) inferred from X-ray observations were used. These are
$a_{\rm o}=24$ kpc for Cygnus A, $a_{\rm o}=140$ kpc for 3C 219 and
$a_{\rm o}=204$ kpc for 3C 215.}
\label{tab:deriv}
\begin{tabular}{cllcccc}
& & & $\theta /{\rm degrees}$ & $Q_{\rm o}/{\rm W}$ & $\rho_{\rm o}/{\rm
kg m}^{-3}$ & $\rho_{\rm o}/{\rm cm}^{-3}$\\
\hline\\
& & 2-D & $71^{+16}_{-29}$ & $6.2^{+2.9}_{-.2} \cdot 10^{38}$ &
$1.7^{+1.9}_{-.3} \cdot 10^{-23}$ & $9.9^{+11}_{-1.7} \cdot 10^{-3}$\\[1.5ex]
& \raisebox{2.25ex}[-2.25ex]{5" resolution} & 1-D & $79^{+10}_{-18}$ & $5.3^{+.3}_{-.7} \cdot 10^{38}$ &
$1.3^{+1.1}_{-.1} \cdot 10^{-23}$ & $7.4^{+6.1}_{-.9} \cdot 10^{-3}$\\[1.5ex]
\raisebox{2.25ex}[-2.25ex]{Cygnus A, eastern lobe}  & & 2-D & $87^{+3}_{-11}$ & $6.5^{+.4}_{-1.3} \cdot 10^{38}$ &
$1.9^{+.6}_{-.3} \cdot 10^{-23}$ & $1.1^{+.4}_{-.2} \cdot 10^{-2}$\\[1.5ex]
& \raisebox{2.25ex}[-2.25ex]{1.3" resolution} & 1-D & $84^{+6}_{-28}$ & $4.2^{+.5}_{-.7} \cdot
10^{38}$ & $1.4^{+1.6}_{-.4} \cdot 10^{-23}$ & $8.4^{+9.4}_{-2.0}
\cdot 10^{-3}$\\[3ex]
& & 2-D & $75^{+14}_{-56}$ & $1.1^{+.6}_{-.4} \cdot 10^{39}$ &
$2.9^{+22}_{-.6} \cdot 10^{-23}$ & $1.7^{+13}_{-.4} \cdot 10^{-2}$ \\[1.5ex]
& \raisebox{2.25ex}[-2.25ex]{5" resolution} & 1-D & $83^{+7}_{-21}$ &
$6.7^{+.5}_{-1.3} \cdot 10^{38}$ & $2.1^{+2.8}_{-.9} \cdot 10^{-23}$
& $1.2^{+1.7}_{-.5} \cdot 10^{-2}$ \\[1.5ex]
\raisebox{2.25ex}[-2.25ex]{Cygnus A, western lobe} & & 2-D & $82^{+2}_{-4}$ & $7.3^{+2.1}_{-2.4} \cdot 10^{38}$ &
$3.6^{+.8}_{-.6} \cdot 10^{-23}$ & $2.1^{+.5}_{-.3} \cdot 10^{-2}$\\[1.5ex]
& \raisebox{2.25ex}[-2.25ex]{1.3" resolution} & 1-D & $79^{+11}_{-25}$ & $5.1^{+.7}_{-1.2} \cdot
10^{38}$ & $1.7^{+2.4}_{-.7} \cdot 10^{-23}$ & $9.8^{+14}_{-4.2}
\cdot 10^{-3}$\\[3ex]
Cygnus A & observed  & & $76$ & \ldots & $2.4 \cdot 10^{-23}$ & $1.4 \cdot
10^{-2}$\\[5ex]
& & 2-D & $59^{+12}_{-6}$ & $1.6^{+.8}_{-.1} \cdot 10^{38}$ & $2.9^{+.4}_{-1.3} \cdot 10^{-25}$ & $1.7^{+.2}_{-.7}
\cdot 10^{-4}$\\[1.5ex]
\raisebox{2.25ex}[-2.25ex]{3C 219} & \raisebox{2.25ex}[-2.25ex]{4.3" resolution} & 1-D & $65^{+24}_{-14}$ & $1.9^{+.1}_{-.3} \cdot
10^{38}$ & $7.4^{+5.3}_{-3.8} \cdot 10^{-25}$ & $4.3^{+3.2}_{-2.2}
\cdot 10^{-4}$\\[1.5ex]
3C 219 & observed & & $32$ & \ldots & $9.2 \cdot 10^{-24}$ & $5.4 \cdot
10^{-3}$\\[5ex]
& & 2-D & $31^{+32}_{-21}$ & $6.7^{+37}_{-3.5}\cdot 10^{38}$ & $1.5^{+3.0}_{-1.2} \cdot
10^{-25}$ & $8.8^{+18}_{-7.3} \cdot
10^{-5}$\\[1.5ex]
\raisebox{2.25ex}[-2.25ex]{3C 215} & \raisebox{2.25ex}[-2.25ex]{1.9" resolution} & 1-D & $43^{+57}_{-5}$ & $3.0^{+.6}_{-.1} \cdot
10^{38}$ & $8.2^{+6.4}_{-7.1} \cdot 10^{-26}$ & $4.8^{+3.8}_{-4.1} \cdot
10^{-5}$\\[3ex]
3C 215 & observed  & & $32$ & \ldots & $9.6 \cdot 10^{-24}$ & $5.6\cdot
10^{-3}$
\\[0.5ex]
\hline
\end{tabular}
\end{center}
\end{table*}

Using Eqs. (\ref{pressure}) and (\ref{length}) The power of the
jet, $Q_{\rm o}$, and the central value of the density distribution of
the gas surrounding Cygnus A, $\rho _{\rm o}$, are calculated. The
results are given in Table \ref{tab:deriv}. Here I assume that the
core radius of the environmental density distribution in Eq.
(\ref{king}) is given by $a_{\rm o}=24$ kpc. The viewing angle
$\theta$ is inferred from the flux ratio of the jet to the counter-jet
in the two lobes of Cygnus A (Hardcastle et
al. 1999\nocite{hapr99}). This assumes that the two jets are identical
and that the observed flux ratio is entirely due to relativistic
beaming effects. Furthermore, a constant bulk velocity within the
jets, $v_{\rm j}$, is assumed and set to 0.62 c. Variations of $v_{\rm
j}$ across the source and asymmetries between the two jet sides will
significantly influence the estimate for $\theta$. However, the model
is consistent with the estimate given by Hardcastle et
al. (1999)\nocite{hapr99}.

The observed central density of source environment given in Table
\ref{tab:deriv} is derived from X-ray observations ROSAT of the hot
gas surrounding Cygnus A Hardcastle \& Worrall
(2000)\nocite{hw00}. For this, the prescription of Birkinshaw \&
Worrall (1993)\nocite{bw93} for the conversion of central surface
brightness to central proton density was used. The core radius,
$a_{\rm o}$ was estimated by Hardcastle \& Worrall (2000)\nocite{hw00}
from the X-ray observations and I use their value, $a_{\rm o}=24$ kpc,
in converting from the density parameter $\rho _{\rm o} a_{\rm
o}^{\beta}$ given by the model to $\rho _{\rm o}$. The value thus
found from the best-fitting model agrees within the error with the
X-ray observations.

\subsubsection{The western lobe}

The western lobe of Cygnus A is covered by 53.5 independent beams
along the jet and 23.2 beams at the widest point perpendicular to the
jet. For the lower resolution maps I find 14.6 and 6.3 beams,
respectively. Similar to the eastern lobe the observed emission does
not extend all the way from the hot spot to the core and so the model
fit is based on a smaller area. From the 2-dimensional maps I find
$R'_{\rm T}=2.3$ for the widest part of the lobe and the length of the
lobe is 73.0" and 69.5" for the low and high resolution case
respectively. The best-fitting models for the two different
resolutions again agree well. The models yield an age for the western
lobe somewhat higher than that of its eastern counterpart (see Table
\ref{tab:res}) which is mainly caused by its greater length. However,
the pressure within the cocoon is remarkably similar in both
lobes. This implies also good agreement between the estimates for the
jet power and the density of the source environment between the two
sides of Cygnus A (Table \ref{tab:deriv}).

\subsection{3C 219}

For 3C 219 at $z=0.1744$ VLA maps at 1.5 GHz in B-array and 4.9 GHz in
C-array were used. The observations were taken in October and December
1998 by Dennett-Thorpe et al. (in preparation) who also performed
standard reduction on the data set using AIPS. The resolution of the
resulting maps is roughly 4.3" and the rms noise is $2.9 \cdot
10^{-4}$ Jy beam$^{-1}$ at 1.5 GHz and $4.4 \cdot 10^{-5}$ Jansky
beam$^{-1}$ at 4.9 GHz. The northern lobe of 3C 219 has a rather
irregular shape and no clear hot spot (Clarke et
al. 1992\nocite{cbbpn92}). This leads to large ambiguities in the
determination of its length or the geometrical parameters needed for
the model presented here. I therefore only used the southern lobe
which has a length of 40.4" and an aspect ratio at its widest point of
$R'_{\rm T}=1.6$. The lobe is covered by 9.4 independent beams along
the jet axis and by 5.9 beams perpendicular to it. The jet and
counter-jet in 3C 219 are unusually bright and so the jet emission was
removed from the maps of the southern lobe. For the 1-dimensional
comparison method, I extracted $S_{\nu} \left( l \right)$ along a line
off-set by 2" to the south of the line connecting the core of the
source with the hot spot of the southern lobe. This avoids
contamination of the surface brightness distribution by the jet
emission. The hot spot in the southern lobe of 3C 219 is somewhat set
back from the edge of the lobe. An aperture with a radius of 4.3",
i.e. the width of the telescope beam, centered on the surface
brightness peak was removed from the maps. For the 1-dimensional
comparison, only values of $S_{\nu} \left( l \right)$ core-wards of
the hot spot were used.

The parameters of the best-fitting model are given in Table
\ref{tab:res}. The uncertainties of the model parameters is comparable
to those found for the two lobes of Cygnus A. The angle to the LOS of
3C 219 is found to be smaller than that of Cygnus A. Since 3C 219 is a
broad line radio galaxy this is in the expected sense, but the value
of $\theta$ predicted by the model is about double that inferred from
the flux ratio of the jet and counter-jet. As mentioned above, the
jets of 3C 219 are unusually bright and this may reflect some enhanced
disruption of the jet flow by turbulence or even a complete restart of
the jets in this source (Clarke \& Burns 1991\nocite{cb91}). The
latter possibility has let Schoenmakers et al. (2000)\nocite{sbrlk99}
to include this source among their examples of Double-Double Radio
Galaxies (DDRG). The morphology of these sources strongly suggests
restarting jets (Kaiser et al. 2000\nocite{ksr00}). The large jet to
counter-jet flux ratio of 3C 219 may therefore be caused by effects
other than relativistic beaming. The best-fitting value of $\theta
=65^{\circ}$ is consistent with orientation-based unification schemes.

The central density of the gas surrounding 3C 219, $\rho _{\rm o}$,
predicted by the model is considerably smaller than that inferred from
X-ray observations. To derive $\rho _{\rm o}$ I used $a_{\rm o}= 141$
kpc (Hardcastle \& Worrall 1999\nocite{hw99}). This discrepancy will
be discussed in Sect. \ref{sec:disc}.

\subsection{3C 215}

This source is a radio-loud quasar at $z=0.411$ with very irregular
morphology (Bridle et al. 1994\nocite{bhlbl94}). I obtained raw
observational data from the VLA archive at 1.5 GHz in A-array and 4.9
GHz in B-array. The 1.5 GHz observations were carried out by Miley in
May 1986 while the 4.9 GHz observations were taken by Hough in
December 1987. Again standard reduction with AIPS was performed on the
data and resulted in two maps with an angular resolution of 1.9". The
maps were restored using a pixel size of 0.3"$\times$0.3" and the rms
noise is $1.6 \cdot 10^{-4}$ Jansky beam$^{-1}$ at 1.5 GHz and $4.0
\cdot 10^{-5}$ Jansky beam$^{-1}$ at 4.9 GHz. The southern half of 3C
215 is very distorted with the jet bending in various places with
enhanced surface brightness (Bridle et al. 1994\nocite{bhlbl94}). This
part of the source is not consistent with a regular FRII-type lobe
morphology and resembles in its outer regions an FRI-type
structure. Therefore no attempt was made to apply the model to the
southern part of the source. The northern lobe is more regular,
however, the hot spot here is weak and the lobe widens considerably
close to the core in a north-eastern direction. The lobe has a length
of 26.6" and its aspect ratio $R'_{\rm T}$ at the point where the
width of the lobe is greatest is 1.2. The hot spot in the northern
lobe is not located at the very edge of the lobe similar to the
southern lobe of 3C 219. An aperture centered on the surface brightness
peak with a radius 3.8'' corresponding to the size of two telescope
beams was removed from the map.  For the 1-dimensional comparison the
surface brightness distribution was extracted along a line off-set by
about 2" to the east from the core-hot spot direction to avoid
emission from the jet. Again only pixels core-wards from the hot spot
were used in the 1-dimensional case. The northern lobe of 3C 215 is
covered by 14.0 independent beams along the jet and 11.7 beams at the
widest point perpendicular to the jet.

Parameters of the best-fitting model and the derived properties of the
environment of 3C 215 are given in Tables \ref{tab:res} and
\ref{tab:deriv}. The uncertainties for the model parameters are
considerably larger for this source than for the two previous ones.
The viewing angle to the jet axis, $\theta$, is smaller than for
Cygnus A or or 3C 219. This is again consistent with the predictions
of unification schemes as 3C 215 is a quasar. The smaller observed
value is again inferred from the flux ratio of the jet and counter-jet
of 3C 215 (Bridle et al. 1994\nocite{bhlbl94}). Similarly to 3C 219
this ratio may be increased in 3C 215 because of the distorted
morphology of its large scale radio structure. The southern jet does
not seem to be embedded in a cocoon and it is therefore unlikely that
the two jets are intrinsically identical. The flux ratio probably
reflects physical processes other than purely relativistic beaming.

For the determination of $\rho _{\rm o}$ I used $a_{\rm o}=204$ kpc
from Hardcastle \& Worrall (1999)\nocite{hw99}. The density of the
gaseous environment of 3C 215 is predicted to be much lower than
inferred from X-ray observations. Discussion of this point is deferred
to Sect. \ref{sec:disc}.

\section{Discussion}
\label{sec:disc}

\subsection{1-D versus 2-D method}

In general it is expected that an increase in the amount of
information available to a given model fitting procedure should
decrease the uncertainty in the determination of the model
parameters. Comparing the results of the 1 and 2-dimensional methods
for the lobes of Cygnus A one would therefore expect that the
parameter uncertainties decrease for the higher resolution maps. This
is indeed the case for the 2-dimensional method, particularly for the
western lobe. However, the 1-dimensional method shows the
opposite. Here the uncertainties are larger for the higher resolution
maps. This is caused by the attempt to fit a very smooth model for the
surface brightness to observational data which shows considerably
greater local variations than the model. The fact that the model is
unable to fit the local surface brightness structure observed was
already noted in the previous Sect.. In the 2-dimensional case the
off-axis parts of the cocoons provide additional information and the
influence of local structure is therefore to some extent averaged out
in the fitting procedure. In other words, the 2-dimensional method is
able to make use of the larger amount of information in higher
resolution maps. The 1-dimensional method is restricted to a cut
through the lobe. Here, the averaging effect of a larger telescope
beam provides for a smoother surface density profile and the model
fits the data better. 

In the case of radio maps of low resolution which do not or only
barely resolve the lobes perpendicular to the jets along most of their
lengths, the 2-dimensional maps add little information to
1-dimensional cuts along the lobes. In these cases the additional
model parameter $p$ in the 2-dimensional method presented here allows
the model to fit the data with a large range of parameter
combinations. The uncertainties of the model parameters are then
larger than for the 1-dimensional method. This effect can be seen for
3C 215 and, to a lesser extent, for the low resolution maps of the
western lobe of Cygnus A. It is of course possible to fix the value of
$p$ for the 2-dimensional method as well but this does not
significantly improve the constraints on the model parameters compared
to the 1-dimensional method. 

Which method is the best to use for a particular set of radio maps
depends on the quality of the maps. If the radio lobes are well
resolved along to the jet axis as well as perpendicular to it then the
2-dimensional methods will provide better constraints. However, it is
computationally expensive. In the case of poorer resolution the
2-dimensional method will not add anything to the results obtained
from a 1-dimensional comparison. For heavily distorted lobe structures
both methods will fail but the 1-dimensional method may still provide
order of magnitude estimates if the lobes are not entirely dominated
by bright localised structure.

\subsection{Determination of viewing angles}

Orientation-based unification schemes attempt to explain radio
galaxies and radio-loud quasars as essentially the same type of
objects albeit viewed at different angles to the jet axis (Barthel
1989\nocite{pb89}). The broad line radio galaxies may then be
identified as the low redshift analoga to radio-loud quasars. These
unification schemes imply that the viewing angle, $\theta$, to the jet
axis is greater than about $45^{\circ}$ for radio galaxies and smaller
for quasars. Tests of this hypothesis are often inconclusive as the
determination of $\theta$ is difficult in practice. 

As described above, the viewing angle $\theta$ may be determined from
radio observations alone. However, since the model depends mostly on
$\sin \theta$, the uncertainties are large particularly for $\theta
\ge 45^{\circ}$. Fortunately, the other model parameters and the
source and environment properties inferred from these do not depend
strongly on $\theta$ and its error in this range. This agrees with the
results of Wan \& Daly (1998)\nocite{wd98} who study the effects of
the viewing angle on the determination of a variety of source
properties in great detail. The source orientation can only be
determined accurately if the 2-dimensional comparison method is used
on radio maps well resolved perpendicular to the jet axis. This agrees
with the findings of Wan \& Daly (1998)\nocite{wd98} who use a
different source model. Despite the large uncertainties in the model
results presented here it is interesting to note that for the three
sources studied the best-fitting values of $\theta$ are smaller for
the quasar (3C 215) and the broad line radio galaxy (3C 219) than for
the radio galaxy (Cygnus A).

\subsection{Environments of FRII sources}

As I showed above, the model parameters of the best-fitting models can
be used to infer the density parameter, $\rho _{\rm o} a_{\rm
o}^{\beta}$, describing the density distribution in the environment of
FRII objects. It is impossible using just this model to separate this
into the central density, $\rho _{\rm o}$, and the core radius,
$a_{\rm o}$. Furthermore, the model is insensitive to the slope,
$\beta$, of the external density profile (see Sect.
\ref{sec:degen}). To infer central densities, values for the core
radius and $\beta$ must be taken from other observations. In Table
\ref{tab:deriv} $\rho _{\rm o}$ for the three example sources is shown
for the core radii given by Hardcastle \& Worrall (1999)\nocite{hw99}
and $\beta =1.5$. As noted in Sect. \ref{sec:appl}, the central
densities found for 3C 219 and 3C 215 are inconsistent with those
derived from the X-ray observations. 

In several studies it was found that the pressure of the FRII source
environment derived from X-ray observations apparently exceeds the
pressure inside the radio lobes (e.g. Hardcastle \& Worrall
2000\nocite{hw00} and references therein). The discrepancy in density
found here is essentially the same phenomenon in the framework of
isothermal density distributions for the source environments as
described by the $\beta$-model, Eq. (\ref{beta}). All recent
models for the evolution FRII sources, including the one discussed
here, are based on the assumption that the cocoon is overpressured
with respect to its surroundings. The X-ray observations seem to
contradict this assumption.

Can the discrepancies be resolved? In the model described here I
approximate the density distribution in the source environment,
assumed to follow the $\beta$-model, by power laws. For cocoons
extending well beyond the core radius the exponent of the power law is
given by $\beta = 3\beta'$. In the case of Cygnus A this is justified
as both lobes extend over more than 100 kpc while $a_{\rm o}=24$
kpc. In the analysis $\beta =1.5$ was used which is identical with the
result of Hardcastle, $\beta' =0.5$. For 3C 219 and 3C 215 $\beta'
=0.9$ from the X-ray observations and in both sources the radio lobes
only extend to about $1.3 a_{\rm o}$. Formally the model cannot be
applied to these two sources because the underlying dynamical model is
valid only for $\beta \le 2$ or $\beta' \le 2/3$. However, as the
cocoons in both sources do not extend much beyond the core radius, the
density distribution in the environments of 3C 219 and 3C 215 are well
represented by a power law with exponent $\beta =1.5$. In these cases
$\rho {\rm o}$ found from the model is not identical to the central
density of the $\beta$-model. From Eq. (\ref{beta}) it can be
seen that the latter is given by $\rho _{\rm o} \left( 1 +x^2
\right)^{3 \beta ' /2}=3.8 \rho _{\rm o}$, where $x$ is the length of
the radio lobe in units of $a_{\rm o}$. This correction is
insufficient to make the model results consistent with the X-ray
observations of 3C 219 and 3C 215.

X-ray observations of the hot gaseous environment of AGN are
influenced by the presence of the active galaxy. Before the
environment properties can be extracted, the bright X-ray emission of
the AGN itself appearing as a point source must be carefully
removed. In the case of radio-loud objects the large scale structure
caused by the jets can also alter the X-ray emission of its
surroundings. The magnitude of these effects is difficult to estimate
if the X-ray observations do not fully resolve the scale of the
cocoon. The hot spots at the end of the jets are strong sources of
inverse Compton scattered X-ray photons. When resolved, these inverse
Compton scattered photons are found to distort the X-ray contours of
the extended emission (Cygnus A, Carilli et al. 1994\nocite{cph94}; 3C
295, Harris et al. 2000\nocite{hnp00}). The more extended cocoon
material itself may also act as a scatterer of CMB photons or of AGN
emission (Brunetti et al. 1999\nocite{bcsf99}). Finally, the bow shock
surrounding the cocoon compresses and heats the gas in the source
environment. Kaiser \& Alexander (1999b)\nocite{ka98b} give an
estimate for the expected X-ray luminosity from this shocked layer of
gas,

\begin{eqnarray}
L_{\rm x} & \sim & 9.1 \cdot 10^{13} \frac{I^2}{a_2} \frac{\left( \rho _o
a_o^{\beta} \right) ^2}{a_1} L_{\rm j}^{\left(7-5\beta \right)/3}
\sqrt{\overline{P/R}} \nonumber \\
& \times & \left[ \exp \left( -4.0 \cdot 10^{-7} a_1^2
L_{\rm j}^{\left(4 -2 \beta \right)/3} \nu / \overline{P/R} \right)
\right] _{\nu _2}^{\nu _1},
\end{eqnarray}

\noindent where two typographical errors are corrected. Expressions
for $I^2/a_2$, $a_1$ and $\overline{P/R}$ may be found in Kaiser \&
Alexander (1999b)\nocite{ka98b}. The square brackets mean the
difference of the exponential function at the limits of the observing
band $\nu _1$ and $\nu_2$ in the rest frame of the source. The
resulting X-ray luminosity can be converted to a ROSAT count rate
using the internet version of PIMMS\footnote{PIMMS was programmed by
K. Mukai at the High Energy Astrophysics Science Archive Research
Center of NASA, available at {\em
http://heasarc.gsfc.nasa.gov/Tools/w3pimms.html}\/}. For the
best-fitting model parameters for Cygnus A this implies that roughly
2\% of all counts of the ROSAT observations presented by Hardcastle \&
Worrall (2000)\nocite{hw00} attributed to the entire extended emission
come from the layer of shocked gas in between bow shock and cocoon of
this source. For 3C 219 this region around the southern lobe alone
contributes about 10\% of the relevant counts. The contribution of the
northern lobe of 3C 215 is negligible ($\sim 0.1$\%). However, in this
case at least three very compact radio emission regions along the
distorted jets are detected which may be powerful inverse Compton
sources (Bridle et al. 1994\nocite{bhlbl94}). Only X-ray observations
resolving the scale of the radio structure in this and other sources
will allow us to determine the contribution to the total X-ray
emission from such compact regions and hot spots.

Even without detailed observations it is clear that powerful radio
sources can contribute significantly to the extended X-ray emission in
the central part of their environments. Helsdon \& Ponman
(2000)\nocite{hp00} show for the case of loose groups of galaxies that
such an overestimate at the centre of an X-ray surface brightness
profile may lead to overestimates of the core radius and also of
$\beta'$ when fitted with a $\beta$-model. Using these overestimated
values of $a_{\rm o}$ and $\beta$ then yields values for $\rho _{\rm
o}$ which are too low compared to the `real' central density. An exact
analysis of the magnitude of this effect is beyond the scope of this
paper. In any case, this effect may explain the discrepancies found
between the two methods of estimating the central densities and
pressures of the FRII environments.

\section{Conclusions}
\label{sec:conc}

In this paper an analytical model for the surface brightness
distribution of the cocoons of FRII-type radio sources is
developed. The model is based on the self-similar model of Kaiser \&
Alexander (1997, KA)\nocite{ka96b} and its extension by Kaiser et
al. (1997, KDA)\nocite{kda97a}. The cocoon is split into small volume
elements, the temporal evolution of which is traced
individually. These elements are then identified with cylindrical
slices which are rotationally symmetric about the jet axis. The bulk
backflow of these slices is determined self-consistently from the
constraints on the cocoon shape. Thus a 3-dimensional model of the
synchrotron emissivity is constructed. Projecting this along the LOS,
the model yields surface brightness profiles in one or two dimensions
in dependence of several model parameters. It is shown that a number
of degeneracies prevent constraints to be placed on all
parameters. Comparatively robust estimates can be found for the cocoon
pressure, $p_{\rm c}$, the source age, $t$, the angle of the jet to
the LOS, $\theta$, and, if the 2-dimensional comparison method is
used, the initial slope of the energy distribution of the relativistic
particles, $p$.

The model may be viewed as an extension of the classical spectral
aging methods. However, it accurately takes into account the loss
history of the relativistic particles and the evolution of the
magnetic field in the cocoon. The age estimates derived from the model
are therefore accurate and not only lower limits. It is also shown
that diffusion of relativistic particles cannot significantly distort
the energy spectrum of the relativistic particles in the cocoon.

In Sect. \ref{sec:appl} the model predictions were compared with
observational data for the three sources Cygnus A, 3C 219 and 3C
215. The four free model parameters are constrained to within at
least a factor 2. The uncertainties of the cocoon pressure are
remarkably small, less than 50\%, while the source ages are
constrained by the model to within 80\% or better. The errors on the
viewing angle are large as the model mainly depends on $\sin \theta$
which changes in the range of orientation angles found here
(31$^{\circ}$ to 90$^{\circ}$) by only 52\%. The uncertainties in the
model parameters translate to errors in the determination of the jet
power, $Q_{\rm o}$, and the density parameter, $\rho _{\rm o} a_{\rm
o}^{\beta}$. However, the resulting uncertainties still allow at least
order of magnitude estimates for these quantities.

The model results do not depend strongly on the resolution of the
radio observations used in the fitting procedure. The results for both
lobes of Cygnus A derived at 1.3" and 5" resolution are consistent
with each other within the errors. However, the uncertainties in the
parameter determination increase significantly with decreasing
resolution. For radio maps which barely resolve the lobes
perpendicular to the jet axis, the 2-dimensional method adds little in
terms of constraints on model parameters to the 1-dimensional
method. From the examples studied here it is not possible to give a
firm lower limit for the resolution of maps used to fit with the
model. However, the parameter uncertainties calculated here suggest
that at least 4 independent beams along the jet axis are required. As
was pointed out in Sect. \ref{sec:back}, distorted lobe structures
may indicate large turbulent flows within the cocoon. As the model is
based on the assumption of largely non-turbulent backflow it cannot be
used for such irregular sources.

The model can be used to infer the viewing angle of FRII-type sources
to the LOS. Despite the large uncertainties in this determination, the
three sources modeled here show the dependence of $\theta$ on the
spectral type of the host galaxies expected from orientation-based
unification schemes.

The central density of the source environments found using this model
is consistent with X-ray observations in the case of Cygnus A. In the
case of 3C 219 and 3C 215 the estimates are too low. These
discrepancies may stem from overestimates from X-ray observations of
the core radius, $a_{\rm o}$, and slope, $\beta$, of the density
distributions. They are caused by the contribution of the large scale
structure, i.e. cocoons and hot spots, of the radio sources to the
X-ray emission. The, at present, insufficient spatial resolution of
X-ray maps prevents the removal of these contributions before the
determination of $a_{\rm o}$ and $\beta$.

The model allows the simultaneous determination of a number of key
parameters of extragalactic radio sources of type FRII. Together with
the 1 and 2--dimensional comparison methods, it is straightforward to
analyse large numbers of objects drawn from complete samples of
extragalactic radio sources. The quality of the radio data needed is
only moderate and will already be available for many of the objects in
question. However, the large scale structure of the objects should be
reasonably regular in the sense that large scale turbulent motion
within the cocoon is unlikely. Within these limits the method can
provide order of magnitude estimates for the properties of the sources
and their environments. This may help to answer some of the questions
related to the cosmological evolution of the FRII population and
unification schemes of radio-loud AGN.

\section*{Acknowledgments}

It is a pleasure to thank P. N. Best, J. Dennett-Thorpe,
H. J. Hardcastle and M. Lacy for valuable discussions. I would also
like to thank P. N. Best and A. P. Schoenmakers for help with the
reduction of the radio data and J. Dennett-Thorpe for providing the
maps of 3C 219. Many thanks to the anonymous referee for valuable
comments.

\appendix

\section{Diffusion of relativistic particles in the cocoon}

\subsection{Quasi-linear anomalous diffusion}

In the model it will be assumed that the magnetic field within the
cocoon is completely tangled on a scale $r_{\rm t}$ much smaller than
the size of the cocoon, $L_{\rm j}$. In this case the relativistic
electrons can diffuse quickly within `patches' of coherent magnetic
field of a size $r_{\rm t}$. However, it is difficult for them to
escape their patch as this would involve diffusion perpendicular to
field lines. In the presence of turbulent motion within the plasma
underlying the magnetic field, patches can be stretched out and in
this case the diffusion of charged particles into an adjacent patch
becomes more likely. Duffy et al. (1995)\nocite{dkgd95} calculate how
far a given electrons has to travel along a field line in its original
patch before the patch becomes so stretched out that it can
escape. Note here that this treatment is valid only in the
quasi-linear regime for which the relative amplitude of irregularities
of the magnetic field is much smaller than the ratio of the turbulent
correlation lengths perpendicular and parallel to the local magnetic
field. The efficiency of this diffusion then depends crucially on
whether the electron is able to travel this distance ballistically or
whether it must diffuse along the field line. The more efficient
ballistic regime requires that

\begin{equation}
r_{\rm g} > \sqrt{\kappa _{\perp} t_{\rm e}}, 
\label{ineql}
\end{equation}

\noindent where $r_{\rm g}$ is the gyro-radius of the particle in the
field, $\kappa _{\perp}$ is the diffusion coefficient perpendicular to
the field and $t_{\rm e}$ is the time it takes the particle to escape
the patch. For a relativistic electron moving at speed $v$ and
corresponding Lorentz factor $\gamma$ following Chuvilgin \& Ptuskin
(1993)\nocite{cp93} $\kappa _{\perp} = \epsilon \kappa _{\rm B} /
\left( \epsilon +1 \right)$ with $\kappa _{\rm B} = \gamma v^2 m_{\rm
e} c / \left( 3 e B \right)$ the Bohm diffusion coefficient and
$\epsilon = \nu _{\rm col} / \nu _{\rm g}$, where $\nu _{\rm col}$ is
the rate of collision of the particle and $\nu _{\rm g}$ is its
gyro-frequency. For the assumption of Blundell \& Rawlings
(2000)\nocite{br00} that the coherence length of the magnetic field is
roughly 10 kpc in all directions, the inequality (\ref{ineql}) yields
$\epsilon < 7 \cdot 10^{-12}$ for a magnetic field strength of 130
$\mu$G, appropriate for the fiducial model of Sect. \ref{sec:degen},
and an electron with a Lorentz factor 1000. This implies that the time
between collisions of this electron must be greater than 2 Myr,
i.e. the mean free path of the electron is of order 600 kpc. This is
clearly unphysical in the case of $L_{\rm j} \sim 100$ kpc as
relativistic particles would then simply escape the cocoon very
quickly.

The relativistic electrons in the cocoon must therefore diffuse along
the magnetic field lines in between jumps from one patch of coherent
magnetic field to another. The expression of Duffy et
al. (1995)\nocite{dkgd95} for the rms diffusion length after a time
$t$, $x$, in this case can be approximated by

\begin{equation}
x \sim \sqrt{- \frac{r_{\rm g} c t}{3 \epsilon \log \left( \sqrt{2}
\epsilon \right)}}.
\end{equation}

\noindent Assuming that the mean free path of the electron is less
than the coherence length of the magnetic field, i.e. less than 10
kpc, I find for the same magnetic field and Lorentz factor $x < 3$ kpc
if $t=1$ Myr. A significant mixing of relativistic particles along the
cocoon due to anomalous diffusion is therefore unlikely.

\subsection{Non-linear diffusion}

Of course, it may be argued that as the above analysis only applies to
the quasi-linear regime, the diffusion timescale in a highly turbulent
flow may be much shorter. Consider such a flow to be present in the
cocoons of FRII-type radio sources. In this case, the relativistic
electrons may diffuse through the observed lobes within a time short
compared to the age of the source. However, since diffusion is a
stochastic process and the geometry of the cocoon is elongated, most
of the particles will leave the cocoon sideways and will be lost to
the surrounding gas before traveling large distances along the
cocoon. It is likely that the diffusion time for the relativistic
particles depends on their energy and so the diffusion losses, if
present, will significantly change the emission spectrum which is not
observed (e.g. Roland et al. 1990\nocite{rhp90}). Even in the case of
efficient energy independent diffusion the observed radio lobes should
show diffuse edges in low frequency radio maps. Again this is not
observed (e.g. Roland et al. 1990\nocite{rhp90}, Blundell et
al. 2000a,b\nocite{brwkp00}\nocite{bkp00}).

From the above it is clear that in the presence of very efficient
diffusion some special confinement mechanism for the relativistic
particles in the cocoon preventing their escape sideways is
needed. This may be provided for by the compression and shearing of
the tangled magnetic field at the edges of the cocoon. This process
will align the magnetic field close to the cocoon edge with this
surface and therefore act as a kind of magnetic bottle. The order thus
introduced in the originally tangled magnetic field due to this
process leads to an enhanced polarisation of the emitted radiation in
this region (e.g. Laing 1980\nocite{rl80}). 

Large volume compression ratios are ruled out as the sound speed in
the cocoon is high (e.g. KA). However, for a conservative estimate
consider a volume compression ratio of 10 in the sheet of compressed
material at the edge of the cocoon of a given source. This already
implies that the maximum theoretical value of polarisation of the
synchrotron emission should be observed at the edge of the radio lobes
(Hughes \& Miller 1991\nocite{hm91}). For adiabatic compression of a
tangled magnetic field the strength of the field increases by a factor
of roughly 4.6. The rate of collisions of a given particle is probably
increased as well as the irregularities in the magnetic field are also
compressed. However, the case of $\nu_ {\rm col}={\rm const.}$
provides for a lower limit of $\epsilon$ and so with $\kappa _{\perp}
= \epsilon \kappa _{\rm B} / \left( \epsilon +1 \right)$ it is clear
that the diffusion coefficient within the compressed region
perpendicular to the field lines decreases by a factor 25 at most
compared to the inner cocoon. This implies that in a given time a
relativistic particle diffuses a 5 times shorter distance in the
compressed boundary layer compared to the inner regions of the
cocoon. To significantly influence the time the particle remains in
the cocoon this compressed layer must at the very least occupy 20\% of
the width of the cocoon which is unlikely. Note, that although the
magnetic field is compressed in the boundary layer and therefore
aligned with the cocoon surface some field lines may still be
perpendicular to this surface. Along these field lines diffusion will
be even faster allowing many electrons to escape the cocoon.

From this I conclude that diffusion, even if it is highly effective in
the inner cocoon and there exists a compressed boundary layer along
the edges of the cocoon, will not alter the distribution of
relativistic particles within the cocoon. This allows us to use the
spatial distribution of the synchrotron radio emission of FRII sources
to infer their age. The model developed in the following can be viewed
as an extension to the classical spectral aging methods in that it
takes into account the evolution of the magnetic field in the lobe

\end{document}